\pdfoutput=1
\documentclass[pre,onecolumn,aps,floats,floatfix,unsortedaddress,superscriptaddress]{revtex4-1}
\usepackage{natbib}
\usepackage[pdftex]{graphicx}
\usepackage{color}
\usepackage{amsfonts}
\usepackage{amssymb,amsmath}
\usepackage{mathrsfs}
\usepackage[english]{babel}
\usepackage{epsfig}
\usepackage{epstopdf}
\usepackage{babelbib}
\usepackage{bm}
\usepackage{tcolorbox}
\usepackage{tabularx}
\usepackage{array}
\usepackage{colortbl}
\tcbuselibrary{skins}
\usepackage{xr-hyper}
\usepackage{hyperref}
\usepackage{xcolor}

\usepackage[minuso]{stmaryrd}
\usepackage[T1]{fontenc}

\newcommand{\tj}[6]{ \begin{pmatrix}
       #1 & #2 & #3 \\
       #4 & #5 & #6 
\end{pmatrix}}

\begin{document}

\title{pH dependence of charge multipole moments in proteins}
\author{An\v{z}e Lo\v{s}dorfer Bo\v{z}i\v{c}}
\email{anze.bozic@ijs.si}
\thanks{Corresponding author}
\affiliation{Department of Theoretical Physics, Jo\v zef Stefan Institute, SI-1000 Ljubljana, Slovenia}
\author{Rudolf Podgornik}
\affiliation{Department of Theoretical Physics, Jo\v zef Stefan Institute, SI-1000 Ljubljana, Slovenia}
\affiliation{Department of Physics, Faculty of Mathematics and Physics, University of Ljubljana, SI-1000 Ljubljana, Slovenia}

\begin{abstract}
Electrostatic interactions play a fundamental role in the structure and function of proteins. Due to ionizable amino acid residues present on the solvent-exposed surfaces of proteins, the protein charge is not constant but varies with the changes in the environment -- most notably, the $pH$ of the surrounding solution. We study the effects of $pH$ on the charge of four globular proteins by expanding their surface charge distributions in terms of multipoles. The detailed representation of the charges on the proteins is in this way replaced by the {\em magnitudes} and {\em orientations} of the multipole moments of varying order. Focusing on the three lowest-order multipoles -- the total charge, dipole, and quadrupole moment -- we show that the value of $pH$ influences not only their magnitudes, but more notably and importantly also the spatial orientation of their principal axes. Our findings imply important consequences for the study of protein-protein interactions and the assembly of both proteinaceous shells and patchy colloids with dissociable charge groups.
\end{abstract}

\maketitle

\section*{INTRODUCTION}

Electrostatic interactions are an important part of the long- and short-range interactions in the biological environment. Their understanding is usually based on the framework of the generalized DLVO (Derjaguin-Landau-Verwey-Overbeek) theory of interactions between colloids in ionic solutions, where the canonical electrostatic and van der Waals components are supplemented by the solvent structure effects of either hydration or hydrophobic type~\citep{Israelachvili2001}. This general decomposition of the interactions remains valid also for proteins in aqueous solution, where it is in addition augmented by short-range recognition and docking interactions as well as specific ion effects~\citep{Leckband1999,Piazza2004}.

While the van der Waals interactions are a functional of the dielectric response function~\citep{Simonson2003,Woods2016}, the electrostatic interactions within and between the proteins are based on their charge distributions. The electrostatic interactions dominate many aspects of protein behavior, and can be modelled on different levels of detail~\citep{Perutz1978,Warshel2006,Gitlin2006}. One of the defining differences between the specific electrostatics of proteins and the generic electrostatics of colloids that needs to be taken into account is the existence of {\em ionizable amino acid residues} in proteins. The interactions and energetics of these residues enable local charge separation, implying a protein-specific distribution of charges~\citep{Tanford1967, Warshel2006} and influencing the protein-protein electrostatic interactions~\citep{Lund2005,Fer2006,Fer2009}. The mechanism of charge separation in turn creates a distinction between the undissociated chargeable groups buried inside the proteins on the one hand and the solvent-exposed and dissociable surface charges on the other~\citep{ProteinPhysics}. This distinction can be blurred, as internal ionizable groups can be to some extent dielectrically shielded even in the strongly hydrophobic protein core~\citep{Freed2008, Moreno2010}. 

Identification of the dissociable, solvent-exposed amino acid residues is the first step in obtaining the description of charge distributions in proteins. Afterwards, one needs to take into account the proper description of the dissociation mechanism for the deprotonated carboxylate of aspartic and glutamic acid, deprotonated hydroxyl of the tyrosine phenyl group, the protonated amine group of arginine and lysine, and the protonated secondary amine of histidine~\citep{Gitlin2006,Jensen2008}. Since the dissociation process and the local electrostatic field are coupled via the charge regulation mechanism~\citep{Markovich2016}, the dissociation rate depends on the local $pH$ that can be obtained only self-consistently~\citep{Szleifer2012, Nap2014}. In principle, only detailed {\em ab initio} simulations can provide a detailed quantification of the partial charges buried inside or exposed on a protein surface. These simulations are, however, usually hampered by the sheer size and number of atoms one needs to invoke in order to achieve a necessary amount of realism for the calculations~\citep{Payne2013,Adhikari2014,Eifler2016}.

A proper quantification of the electrostatic interactions in proteins requires the encoding of not only the magnitudes of the charges but also of their distribution in space~\citep{Gramada2006,Hoppe2013}. The latter can be represented to any desirable accuracy by a multipole expansion of the charge density, where each term in the multipole series describes a deviation with a specific symmetry from the zeroth-order, spherically symmetric distribution~\citep{Diederich2005}. There are many variants of the multipole expansion, formulated either on the level of amino acids or on the level of complete proteins~\citep{Yuan2014-1,Yuan2014-2,Fletcher2016}, with the most straightforward being the {\em irreducible spherical representation} of the multipoles~\citep{Gramada2006}. This representation is obtained by mapping the charge distribution on the original solvent-accessible protein surface onto a sphere circumscribed to the protein~\citep{Hoppe2013,Postarnakevich2009,Arzensek2015}.

The multipole expansion provides a bridge between a coarse-grained description of the charge density and its detailed microscopic description, the level of detail depending on the multipole order of the expansion used. As such, even multipoles of lower order can provide a {\em signature} of charge distributions in molecules~\citep{Gramada2006,Platt1996,Nakamura1985,Kim2006}, and a small number of higher-order multipoles can account for almost all of the electrostatic field in the aqueous solvent~\citep{Hoppe2013}. In the presence of ionic screening this otherwise standard result is modified, as the effects of charge anisotropy and higher-order multipole moments then extend to the far-field region. In contrast to the standard multipole expansion, the screened electrostatic potential retains the full directional dependence of all multipole moments, an important difference which is often overlooked~\citep{Trizac2000,Kjellander2008,ALB2013a,Kjellander2016}.

With a few exceptions~\citep{Hoppe2013}, the $pH$ dependence of charge distributions in proteins has been standardly studied mostly on the level of the spherically symmetric total charge, that is, the zeroth multipole moment. This approach, however, completely neglects the orientational dependence of the interactions. Due to the peculiarity of the multipole expansion for screened electrostatic interactions, the orientational effects are present even at the lowest multipolar order and can thus -- contrary to the existing analytical models -- never really be ignored~\citep{Trizac2000,Kjellander2008,ALB2013a, Kjellander2016}. In addition, for higher-order multipoles the two components of the multipole expansion, the magnitude of the multipole moments and their spatial distribution, can hardly be separated. We thus investigate not only the $pH$ dependence of the {\em magnitudes} but also of the {\em directions of the principal axes} of the charge dipole and quadrupole moments in proteins, a phenomenon missed also in cases where the higher multipole terms were included~\citep{Li2015,Mcmanus2016,Grant2001,Matthew1985}. It is in fact the orientational dependencies that are of particular importance for the local electrostatic interactions in numerous contexts, even when the coupling between $pH$, protonation states, and protein conformation is not taken into account. The results of our model should thus be relevant for the studies of interaction and assembly in ordered protein assemblies (such as a proteinaceous virus shell~\citep{Siber2012}), between charged Janus colloids~\citep{Hieronimus2016}, in the general context of patchy globular proteins, colloids, and polyelectrolytes~\citep{Yigit2015a,Yigit2015b,Yigit2017,Bianchi2011,Bianchi2014,Stipsitz2015}, or possibly as a driving mechanism for local packing symmetry transitions in the proteinaceous capsid engineering in the presence of supercharged proteins~\citep{Hilvert}.

\section*{MATERIALS AND METHODS}

\subsection*{Protein dataset}

For our study we chose four protein structures from the RCSB Protein Data Bank (PDB)~\citep{PDB}: hen egg-white lysozyme (2lyz), human serum albumin (1e7h), bovine $\beta$-lactoglobulin (2blg), and phage MS2 capsid protein (2ms2, subunit A). These proteins are biologically well-studied, with known structure and role. As such, they are often used to examine the role of electrostatic contributions in protein systems and to study protein assembly and aggregation, where higher-order charge multipoles play a role~\citep{Hoppe2013,Nap2014,Lund2005,Yigit2017,Cardinaux2007}. $\beta$-lactoglobulin, for instance, is known to form dimers or tetramers depending on the $pH$. The capsid protein of MS2 also first assembles into trimers and from there into full capsids made of 180 proteins, imparting to the capsids a $pH$-dependent stability. As our interest lies in exploring the $pH$ dependence of multipole moments and their orientations that could relate to higher-order assembly, we study here only the monomer forms of each protein. In addition, the proteins in our dataset have globular geometry, easily approximated by a sphere, and most of them are fairly small, with lysozyme and MS2 capsid protein being composed of 129 amino acids (AA), and $\beta$-lactoglobulin of 162 AA. Human serum albumin (HSA) is the largest of the four, consisting of 585 AA. While we will focus on the four proteins in the dataset to study the general properties of our model, we note that our approach can also be straightforwardly implemented for any globular protein.

\begin{figure}[!htp]
\begin{center}
\includegraphics[width=0.3\textwidth]{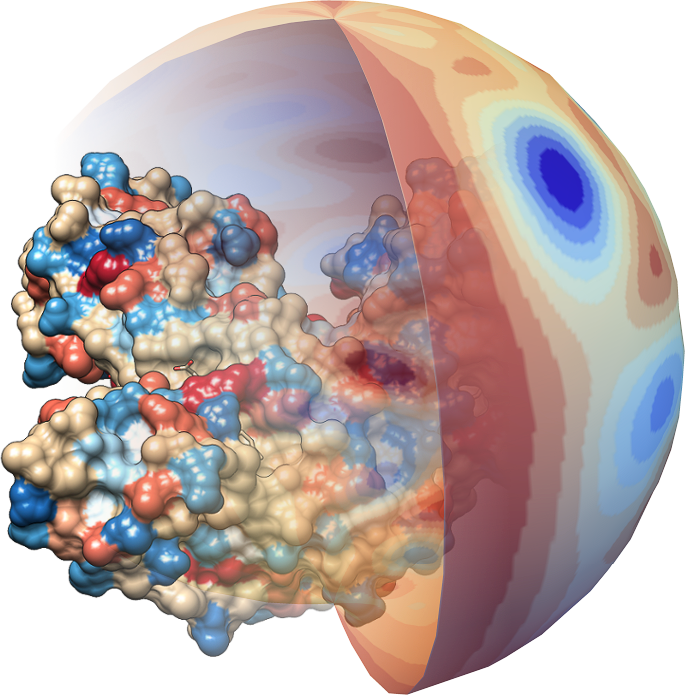}
\end{center}
\caption{Sketch of the model showing a rendering of the surface structure of human serum albumin (1e7h), superimposed onto a circumscribed sphere with projected multipole expansion of the surface charge distribution (up to $\ell_{\max}=12$). AA residues that are charged at $pH=7$ are highlighted in the structure, with colors pertaining to the red spectrum indicating positive charges, and colors in the blue spectrum indicating negative charges. The same color scheme applies to the projection of the charge distribution onto the sphere. The protein structure was rendered with UCSF Chimera~\citep{CHIMERA}.
\label{fig:1}}
\end{figure}

To obtain the surface charges on each protein at a given $pH$ value, we first determine which AA residues are solvent-accessible, and compute the charge on them by using the canonical static dissociation constant $pK_a$ value pertaining to each AA type. By projecting the positions of the charged residues onto a sphere, we obtain the surface charge distribution of each protein, which we then use to compute the electrostatic multipole moments. Details of each of these steps are laid out in the following Subsections, and a sketch of the model is shown in Fig.~\ref{fig:1}.

\subsection*{Relative solvent accessibility of amino acid residues}

Relative solvent accessibility (RSA) of an amino acid indicates its degree of burial in the three-dimensional protein structure, and is important in determining which AA residues can contribute to the surface charge of the protein. We obtain the RSA of each AA in a protein with the help of the UCSF Chimera software~\citep{CHIMERA,MSMS} by normalizing the solvent-exposed surface area of each residue in the protein structure with the surface area of the same type of residue in a reference state~\citep{SESAvalues}. The classification of AA residues as ``buried'' or ``exposed'' is then done on the basis of an RSA cut-off $c$, which typically ranges between 5\% and 30\%~\citep{Pollastri2002,Yuan2002,Chen2005,Wu2017}. We opt for a cut-off of $c=0.25$, thus defining as {\em exposed} all amino acid residues $\mathrm{AA}_k$ with an RSA value greater than 25\%:
\begin{equation}
\mathrm{RSA}(\mathrm{AA}_k)\geqslant0.25\Rightarrow\mathrm{AA}_k\in\mathrm{exposed}.
\end{equation}

The choice of this (arbitrary) cut-off influences the number of charged residues that will be taken into account in our calculation of the surface charge of the protein. How the number of charges varies with the selection of the RSA cut-off is shown in Table~\ref{tab:S1} in the Supporting Material. Similar variations in the number of surface charges could also occur, for instance, due to a different choice of the normalization values (reference states) in the calculation of the RSA~\citep{Tien2013}.

For consistency, we use the RSA cut-off of $c=0.25$ throughout the paper, unless specified otherwise. However, we also test our predictions for several other values of the cut-off and show that it bears no influence on the conclusions obtained in our work.

\subsection*{pH dependence of charged amino acid residues}

Once we know which AAs are solvent-exposed and thus dissociable, we can determine their charge at a given $pH$ value. The charged residues we consider are the aspartic acid (ASP), glutamic acid (GLU), tyrosine (TYR), arginine (ARG), lysine (LYS), and histidine (HIS), while the case of cysteine (CYS) protonation is considered separately in the Discussion and Supporting Material. To obtain the charge on each of the residues as a function of $pH$ we use the acid-base dissociation constants $pK_a$ from Ref.~\citep{CRC} (and listed in Table~\ref{tab:S2} in the Supporting Material). The degree of dissociation of each amino acid AA$_k$ as a function of its $pK_a^{(k)}$ and $pH$ is then given by the Henderson-Hasselbach equation:
\begin{equation}
\label{eq:hh}
q^{\pm}_k=\frac{\pm1}{1+e^{\pm\ln10(pH-pK_a^{(k)})}}
\end{equation}
for bases ($q^{+}_k>0$) and acids ($q^{-}_k<0$), respectively. We note that eq.~\eqref{eq:hh} can be expanded to include the local electrostatic potential at the position of each charge, $\psi(\mathbf{r}_k)$, inducing a $pK_a$ shift~\citep{Ninham1971,Krishnan2017}:
\begin{equation}
\label{eq:pn}
q^{\pm}_k=\frac{\pm1}{1+e^{\pm\ln10(pH-pK_a^{(k)})\mp\beta e_0\psi(\mathbf{r}_k)}},
\end{equation}
where $\beta=1/k_BT$, $T$ is the room temperature, $k_B$ the Boltzmann constant, and $e_0$ is the elementary charge. The $pK_a$ shift due to local electrostatic potential is hard to neglect, especially at low salt concentrations. However, reformulating the Tanford-Kirkwood model~\citep{TanfordKirkwood,Fer2001} by using a linearized Debye-H\"{u}ckel (DH) theory on a low dielectric constant sphere with charge regulation boundary condition, we show that for large (physiological) salt concentrations the effect of the $pK_a$ shift due to the electrostatic potential is relatively small and eq.~\eqref{eq:hh} can be used. Details of the DH model are given in the Supporting Material [eqs.~\eqref{eq:chr}-\eqref{eq:fin}], and its consequences commented on in the Discussion.

Electrostatic effects are not the only possible factor inducing a local $pK_a$ shift of a given amino acid in a protein, as hydrogen bonding and desolvation effects often play an important role. Numerous methods exist for the prediction of $pK_a$ values on different levels of detail, ranging from various molecular dynamics models and {\em ab initio} quantum mechanical approaches to empirical models, which trade the detailed description of a system for a very fast computational time~\citep{Alexov2011}. In order to further compare our results, obtained by assigning the same $pK_a$ value to each amino acid residue group, with a more realistic model including site-site interactions, we use the PROPKA3.1 software~\citep{Sondergaard2011} to predict local $pK_a$ values of each amino acid residue. PROPKA is a widely-used empirical software which uses three-dimensional structure of proteins to estimate desolvation effects and intraprotein interactions based on the positions and chemical nature of the groups proximate to the $pK_a$ sites. The $pK_a$ values predicted by PROPKA do not include salt effects as an input parameter, and thus provide another layer of contrast to the $pK_a$ values shifted due to electrostatic effects, described above. The results obtained in this way are presented in the Discussion and the Supporting Material, where we show that for the purposes of our study, the simple model that we use fares well even when compared to a more realistic model.

\subsection*{Surface charge distribution and electrostatic multipoles}

With the approach outlined in the previous Subsections we can obtain, at any value of $pH$, the positions of charged residues for a given protein, $\bm{r}_k=(x_k,y_k,z_k)$, and the (fractional) charge they carry, $q_k$. To obtain a surface charge distribution we then project them onto a spherical surface, $\bm{r}_k=(R,\Omega_k)=(R,\vartheta_k,\varphi_k)$, so that their positions are characterized only by their solid angle, $\Omega_k$. Here, the radius of the projecting sphere, $R$, can be any characteristic dimension of the protein, its circumscribed radius most often being used for this purpose~\citep{Hoppe2013}. The circumscribed radii of the studied proteins are given in Table~\ref{tab:S3} in the Supporting Material, and fall in the range of $R\sim1$-$2$ nm.

The surface charge distribution of the discrete charges can then be written simply as
\begin{eqnarray}
\label{eq:1}
\nonumber\sigma(\Omega)&=&\frac{e_0}{R^2}\sum_{k\in\mathrm{AA}}q_k\delta(\Omega-\Omega_k)\\
&=&\frac{e_0}{R^2}\sum_{l=0}^\infty\sum_{m=-l}^l\sigma_{lm}Y_{lm}(\Omega),
\end{eqnarray}
if rewritten in the form of an expansion in terms of the irreducible spherical representation of multipoles; $\delta(x)$ is the Dirac delta function and $Y_{lm}(\Omega)$ are the spherical harmonics. From eq.~\eqref{eq:1} we can then obtain the {\em multipole expansion coefficients} $\sigma_{lm}$ as
\begin{equation}
\sigma_{lm}=\sum_{k\in\mathrm{AA}}q_kY_{lm}^*(\Omega_k).
\end{equation}
Contrary to the case of an unscreened Coulomb potential, the Debye screening limits the infinite sum over $l$ in the multipole expansion [eq.~\eqref{eq:1}] to only the first several terms, leading to a coarse grained description as the details of the charge distribution are smeared below the Debye screening length cut-off. Therefore, even in the presence of higher order symmetries (e.g., octahedral or icosahedral), the lower-order multipoles up to and including the quadrupole provide a good signature of the charge distributions in various molecules, accounting for most of the electrostatic field~\citep{Gramada2006, Platt1996, Nakamura1985, Kim2006, Hoppe2013}.

In view of this, and limiting ourselves to the limit of strong screening, we will focus on the three lowest-order multipoles, the total charge (monopole with rank $\ell=0$), the dipole moment ($\ell=1$), and the quadrupole moment ($\ell=2$). The total charge is independent of the choice of coordinates and can be obtained simply as
\begin{equation}
q=\sum_{k\in\mathrm{AA}}q_k,
\end{equation}
where $k$ runs over all of the charged amino acids in the protein. The total charge as well as the higher-order multipoles will be given in units of elementary charge, $e_0$, unless noted otherwise.

For simplicity, in dealing with the dipole and quadrupole moment we will also rescale all our calculations with the characteristic radius of the protein $R$, obtaining an orientational distribution on a unit sphere. The positions of the charges become unit vectors $\bm{n}_k$, expressed in spherical coordinates as $\bm{n}_k=(1,\vartheta_k,\varphi_k)$. As we will also be interested in the orientation of the dipole vector and the eigenvectors of the quadrupole tensor, it will be easier for us to deal with them in Cartesian coordinates. The expressions for the dipole vector $\bm{\mu}$ and the quadrupole tensor $\mathcal{Q}$ are then~\citep{GrayGubbins}
\begin{eqnarray}
\label{eq:2}\bm{\mu}&=&\sum_{k\in\mathrm{AA}}q_k\bm{n}_k,\\
\label{eq:3}\mathcal{Q}&=&\frac{1}{2}\sum_{k\in\mathrm{AA}}q_k\left(3\bm{n}_{k}\bm{n}_{k}-\bm{1}\right),
\end{eqnarray}
or in an obvious component notation ($\mu_i$ and $Q_{ij}$) that we do not write down explicitly. The proper units for the multipoles can be obtained by multiplying the expressions in eqs.~\eqref{eq:2} and~\eqref{eq:3} with the corresponding power of the characteristic protein radius $R^\ell$ (i.e., $\ell=1$ for the dipole and $\ell=2$ for the quadrupole). Cartesian components of both dipole and quadrupole moments can also be very easily transformed into a spherical form in which they might be more suitable for analytical or numerical calculations (cf.\ Ref.~\citep{GrayGubbins}).

We choose the center-of-mass of each protein for the origin of the coordinate system in which we compute the surface charge distribution and the electrostatic multipoles. Since the monopole moment (the total charge) of the surface charge distribution is always non-zero except at the isoelectric point, the higher multipoles are dependent on the choice of the origin. However, the transformations of the dipole and quadrupole moment to other coordinate systems are very simple and given in Ref.~\citep{GrayGubbins}.

\subsection*{Dipole and quadrupole principal axes}

Surface charge distributions on the proteins, and therefore their multipole expansions, will change with $pH$. This will influence not only the magnitudes of the individual multipoles but also their orientation in space. Consequently, we will be interested in the orientations of the dipole and quadrupole distributions with respect to the original (reference) coordinate system. For each protein, this coordinate system is derived from its PDB entry; the exact orientation of the original coordinate system with respect to the protein structure will not be of interest to us, since we will be interested in relative changes of the dipole and quadrupole orientations with $pH$. While the monopole moment is rotationally invariant, we can always find a rotation of the original coordinate system by keeping the protein structure fixed so that either {\em (i)} the dipole vector has a non-zero component only in the $z$ direction, or {\em (ii)} the quadrupole tensor is diagonal, with the largest eigenvalue aligned along the $z$ axis. We will refer to these two $z$ axes in the rotated coordinate systems as the {\em dipole and quadrupole principal $z$ axes}, respectively.

The dipole moment, being a vector, has three independent components: in the original coordinate system they are the $\mu_{x'}$, $\mu_{y'}$, and $\mu_{z'}$. Upon rotation into the dipole coordinate system, the dipole moment has a non-zero component only along its principal $z$ axis, $\bm{\mu}=(0,0,\mu_z)$, where $\mu_z=(\mu_{x'}^2+\mu_{y'}^2+\mu_{z'}^2)^{1/2}$. The two remaining parameters are now the two angles needed to align the original $z'$ axis into the new, principal $z$ axis. (Since the only non-zero component is in the $z$ direction, the position of the new $x$ and $y$ axes is irrelevant.)

The quadrupole moment is a symmetric tensor, thus having five independent components. In the quadrupole coordinate system specified by its eigenvectors the tensor becomes diagonal, and we can always order its eigenvalues by value so that the largest one is oriented along the principal $z$ axis, $Q_{xx}\leqslant Q_{yy}\leqslant Q_{zz}$. Since the quadrupole tensor is traceless, we can also express one of its eigenvalues with the other two, e.g., $Q_{yy}=-(Q_{xx}+Q_{zz})$. This leaves us with three more independent components which we can identify with, for instance, the three Euler angles needed to rotate the original coordinate system into the one defined by the quadrupole eigenvectors. Due to the symmetry of the quadrupole, we will also restrict the location of its principal $z$ axis only to the upper hemisphere of the circumscribed sphere.

\subsection*{Quadrupole ratio}

In the case of the quadrupole moment we will also be interested in the projections of the surface charge distributions along the $y$ and $x$ axes in the quadrupole coordinate system. Because of our choice of the ordering of the eigenvalues, $Q_{zz}$ will always take on the largest positive value and $Q_{xx}$ the largest negative value. Thus, a ratio of these two eigenvalues, $|Q_{xx}/Q_{zz}|$, will provide us with information on what proportion of the quadrupole distribution is aligned with the principal $z$ and $x$ axis, respectively. (The $y$ axis information can be again omitted due to $\mathrm{Tr}\mathcal{Q}=0$.)

When the quadrupole ratio is close to $0.5$, the quadrupole distribution is axial and oriented predominantly along the $z$ axis. Likewise, when the ratio is close to $2$, the distribution is axial but with an opposite sign and oriented along the $x$ axis. On the other hand, when this ratio is close to $1$, the distribution is represented symmetrically in the $z$ and $x$ directions while vanishing along the $y$ axis. In this scenario, the distribution is not axial but corresponds better to a planar one. The ratio of the two quadrupole eigenvalues thus gives us an insight into the spatial distribution of the quadrupole, which can have an impact on, for instance, the interaction and assembly of molecules with a pronounced quadrupole moment. Some examples of the relation between the quadrupole ratio and the spatial distribution of the quadrupole moment are shown in Figs.~\ref{fig:S1} and~\ref{fig:S2} in the Supporting Material.

\section*{RESULTS}\label{sec:results}

\subsection*{pH dependence of multipole magnitudes}

\begin{figure*}[!t]
\begin{center}
\includegraphics[width=0.48\textwidth]{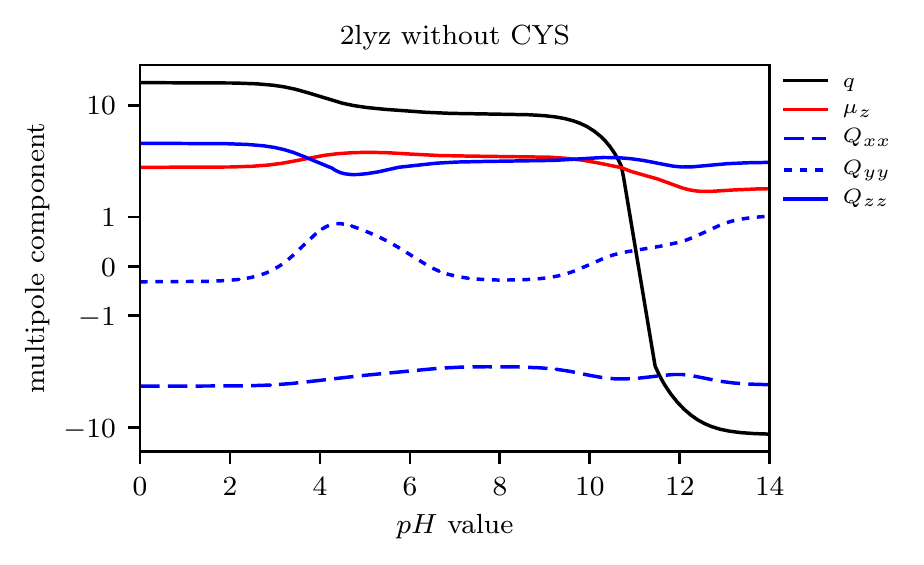}
\includegraphics[width=0.48\textwidth]{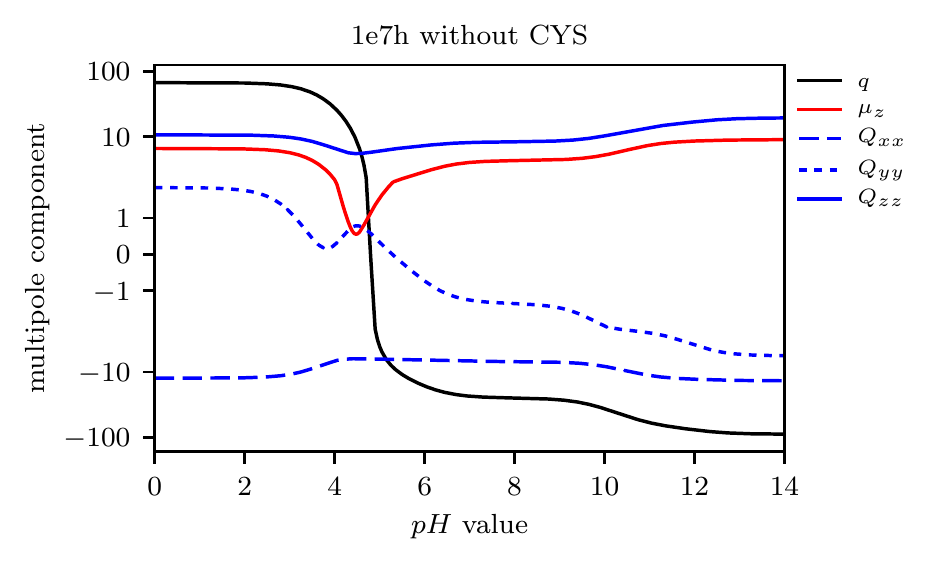}
\end{center}
\caption{Magnitudes of the monopole ($q$), dipole ($\mu_z$), and quadrupole ($Q_{ii}$) components of the surface charge distributions of lysozyme (2lyz) and human serum albumin (1e7h), shown as a function of $pH$. Cysteine acidity is not considered.
\label{fig:2}}
\end{figure*}

Figure~\ref{fig:2} shows the $pH$ dependence of the first three multipole components in the cases of lysozyme and HSA; similar plots for the MS2 capsid protein and $\beta$-lactoglobulin are shown in Fig.~\ref{fig:S4} in the Supporting Material. The total charge on the proteins standardly decreases from positive to negative as the $pH$ increases, crossing the point of zero charge at the isoelectric point, $pI$. The isoelectric points of the proteins are interesting since the monopole moment vanishes and higher-order multipoles take on a greater importance. When cysteine acidity is not considered, the predicted isoelectric point of lysozyme is $pI=11.08$, an alkaline $pI$ that is in very good agreement with the values found in the literature ($pI\gtrsim11$~\citep{Wetter1951,Strang1984}), even though we used a fairly simple method to obtain it. The $pI$ values we obtain for other proteins are listed in Table~\ref{tab:1} and also correspond well with the experimental values: $pI\sim4.7$--$5.6$ for HSA~\citep{Evenson1978,Langer2003} and $pI\sim5.1$ for $\beta$-lactoglobulin~\citep{Majhi2006,Mercadante2012}. The $pI$ value obtained for the phage MS2 capsid protein at a first glance disagrees with the very low reported value of the phage MS2, $pI\sim2.2$--$3.9$~\citep{Michen2010,Dika2015}. The notable discrepancy stems most probably from the presence of the genome in the interior of the virion and the permeability of the capsid to external flow~\citep{Dika2011} -- in studies where the capsid protein alone was considered, the obtained isoelectric point was at a $pI>8$~\citep{Langlet2008, Penrod1996, Schaldach2006}. Our results thus match the ones found in the literature, as we consider only a single capsid protein and not the formed virion (with or without the genome), even though they do not match the experimental results for the formed phage.

In contrast to the total charge, the dipole moment does not change its sign, which is expected, as we always orient the dipole coordinate system so that it points in the $z$ direction. The $pH$ dependence of the dipole magnitude is nonetheless non-monotonic. In a similar fashion, the quadrupole components vary non-monotonically across the range of $pH$ values, with the $Q_{zz}$ and $Q_{xx}$ values always being positive and negative, respectively, due to our definition of their direction. The value of $Q_{yy}$, as a consequence of $\mathrm{Tr}\mathcal{Q}=0$, crosses zero when the $Q_{xx}$ component becomes larger than the $Q_{zz}$ component, or vice versa.

While we use the normalized multipole moments (in units of $[e_0/R^{\ell}]$ in the figures to allow for an easier comparison, we also provide in Table~\ref{tab:1} the multipole magnitudes in the more intuitive units of $[e_0\times\mathrm{nm}^{\ell}]$. We list the dipole and quadrupole magnitudes at the isoelectric point of each protein, as well as the multipole magnitudes at neutral $pH=7$. We can observe a significant decrease in the dipole moment of HSA at its isoelectric point when compared to the dipole moment at neutral $pH$. A closer inspection of Fig.~\ref{fig:2} indeed shows that the dipole moment of HSA exhibits a minimum close to its isoelectric point. HSA also attains by far the largest charge and quadrupole moment of the four proteins at neutral $pH$; however, its dipole moment is comparable to that of the others. Another observation we can draw from Table~\ref{tab:1} is that dipole moments seem to play a bigger role in the cases of $\beta$-lactoglobulin and phage MS2 capsid protein when compared to their quadrupole moments, while something of the opposite is true for HSA and in a lesser way for lysozyme.

\begin{table*}[!htb]
\begin{center}
\begin{tabular}{lcccccccc}
PDB & CYS & $pI$ & $\mu_I$ $[e_0\times\textrm{nm}]$ & $Q_I$ $[e_0\times\textrm{nm}^2]$ & $q_n$ $[e_0]$ & $\mu_n$ $[e_0\times\textrm{nm}]$ & $Q_n$ $[e_0\times\textrm{nm}^2]$ & $\langle|Q_{xx}/Q_{zz}|\rangle_{pH}$ \\
\hline
\hline
2lyz & no & $11.08$ & $1.85$ & $3.86$ & $8.10$ & $2.67$ & $2.92$ & $1.09$ \\
1e7h & no & $4.78$ & $2.27$ & $34.68$ & $-23.45$ & $7.86$ & $38.01$ & $0.93$ \\
2blg & no & $4.47$ & $5.77$ & $2.01$ & $-7.78$ & $7.43$ & $2.64$ & $0.91$ \\
2ms2 & no & $9.78$ & $4.88$ & $2.34$ & $2.00$ & $4.91$ & $2.78$ & $1.02$ \\
\end{tabular}
\end{center}
\caption{Summary of results for the $pH$ dependence of the magnitudes of multipole components when cysteine acidity is not considered. Listed are the isoelectric point, $pI$, where the monopole moment (total charge) vanishes, and the magnitudes of the dipole and quadrupole moment in this point (subscript $I$). We also list the magnitudes of the monopole, dipole, and quadrupole moments at neutral $pH=7$ (subscript $n$). Lastly, we list the quadrupole ratio $\langle|Q_{xx}/Q_{zz}|\rangle_{pH}$, averaged across the entire range of $pH$ values. The magnitudes of the multipole moments correspond to total charge in the case of the monopole, Cartesian norm in the case of the dipole, and $Q=[(Q_{xx}-Q_{yy})^2+Q_{zz}^2]^{1/2}$ in the case of the quadrupole.
\label{tab:1}}
\end{table*}

\subsubsection*{Quadrupole ratio: axial or planar distribution}

\begin{figure}[!t]
\begin{center}
\includegraphics[width=0.48\textwidth]{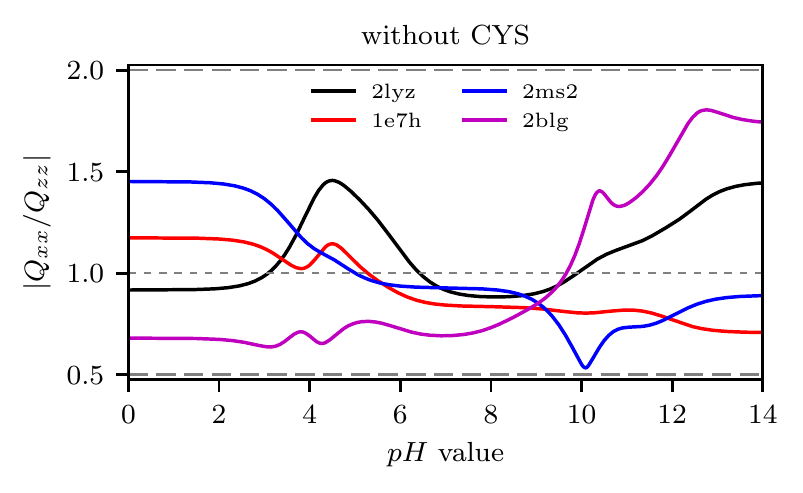}
\end{center}
\caption{Quadrupole ratio $|Q_{xx}/Q_{zz}|$ as a function of $pH$ for all the proteins studied; cysteine acidity is not considered. The ratio determines whether the quadrupole distribution is axial ($|Q_{xx}/Q_{zz}|=0.5$ and $|Q_{xx}/Q_{zz}|=2$), planar ($|Q_{xx}/Q_{zz}|=1$), or somewhere in-between. For additional geometrical interpretation of the quadrupole ratio see Figs.~\ref{fig:S1} and~\ref{fig:S2}.
\label{fig:3}}
\end{figure}

We have seen that both $Q_{zz}$ and $Q_{xx}$ components of the quadrupole tensor exhibit non-monotonic variation with $pH$. To be able to interpret their relationship more easily we also plot their ratio, $|Q_{xx}/Q_{zz}|$, as a function of $pH$ (Fig.~\ref{fig:3}). When either of the two components dominates, we have $|Q_{xx}/Q_{zz}|\sim0.5$ or $|Q_{xx}/Q_{zz}|\sim2$, and the distribution is axial along the $z$ or $x$ axis, respectively. On the other hand, when the two components are comparable, we have $|Q_{xx}/Q_{zz}|\sim1$, and the distribution is ``planar'' in the $x$-$z$ plane. In all the cases studied the quadrupole ratio varies quite a lot, although for most proteins (with the exception of $\beta$-lactoglobulin) it never reaches a fully axial distribution. Conversely, in certain ranges of $pH$ the quadrupole distributions of all the proteins are approximately planar. This is also mirrored by the average values of their quadrupole ratios, which are indeed close to 1 (Table~\ref{tab:1}).

These results indicate that the nature of quadrupole distributions in the proteins used in our work can be influenced by changing the $pH$: At certain values of $pH$, the quadrupole distributions are axial and thus oriented along a single axis (here, due to our definitions, along the $z$ or $x$ axis). At other values of $pH$, however, the quadrupole distribution becomes planar and symmetrically distributed in the $x$-$z$ plane. The nature of the quadrupole distribution can thus change quite drastically with $pH$ even without any concomitant changes in the conformation of the protein~\citep{OBrien2011}, which should influence, for instance, the interaction properties of these proteins.

\subsection*{pH dependence of dipole and quadrupole principal axes}

\begin{figure*}[!b]
\begin{center}
\includegraphics[width=0.8\textwidth]{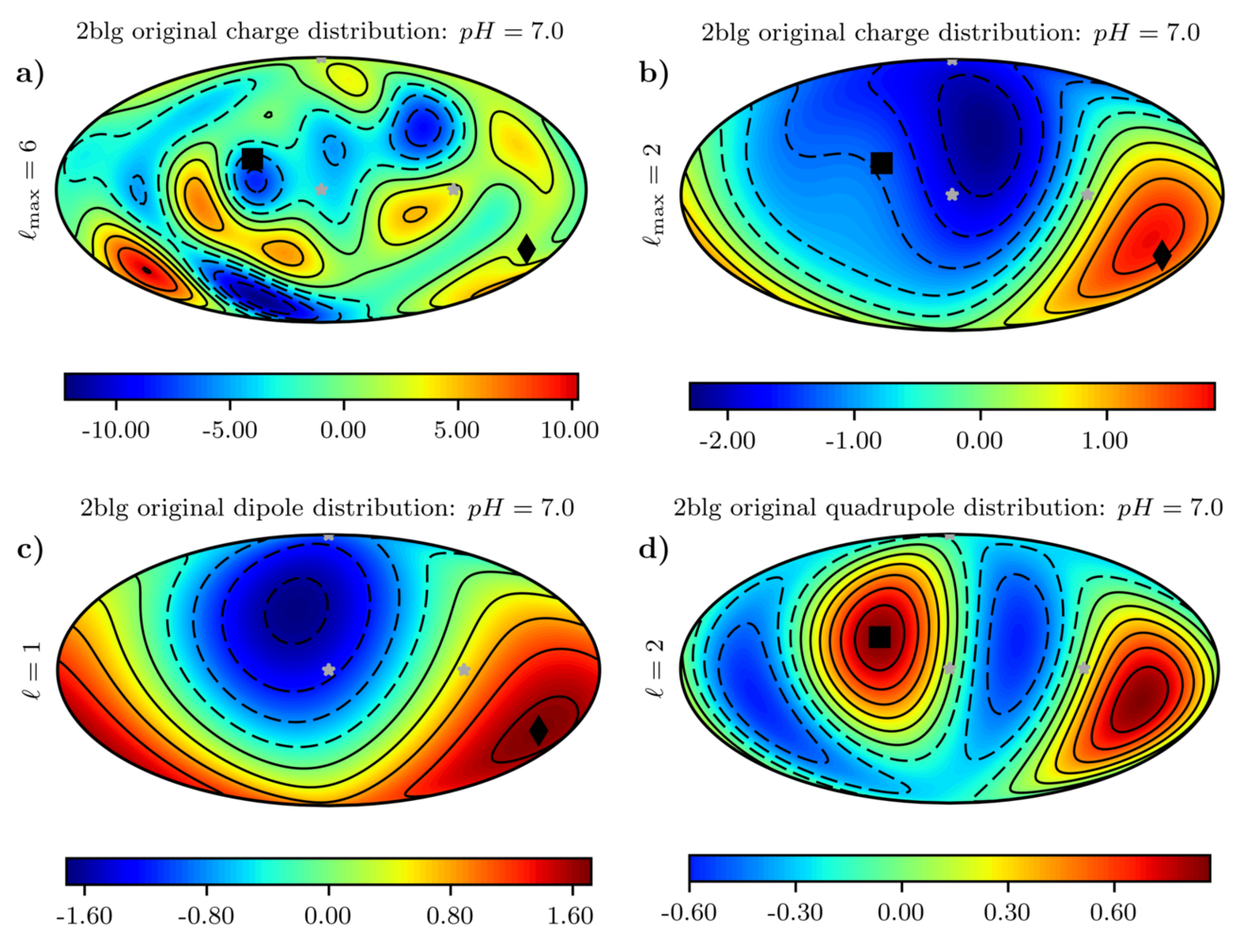}
\end{center}
\caption{Multipole expansion of the surface charge distribution of $\beta$-lactoglobulin (2blg) in the original coordinate system up to {\bf (a)} $\ell_{\max}=6$ and {\bf (b)} $\ell_{\max}=2$. Shown are also the {\bf (c)} dipole and {\bf (d)} quadrupole distributions in the original (reference) system. The distributions are mapped from a sphere to a plane using the Mollweide projection. Black diamonds show the orientation of the $z$ axis of the dipole, and black squares the orientation of the $z$ axis of the quadrupole. Gray stars show the coordinate axes of the original coordinate system. Cysteine acidity is not considered, and all the plots are drawn at $pH=7.0$.
\label{fig:4}}
\end{figure*}

Now that we have seen how the magnitudes of different multipole moments change with $pH$, we turn our attention to the $pH$ dependence of the orientations of the dipole and quadrupole principal axes when compared to the original (reference) coordinate system. In order to indicate the level of detail described by different multipoles, we show in Figs.~\ref{fig:4}a and~\ref{fig:4}b the multipole expansions of the surface charge distribution of $\beta$-lactoglobulin on a fine-grained level with maximum rank of $\ell=6$ and on a coarse-grained level with rank $\ell=2$, respectively. The distributions are mapped from a sphere to a plane using the Mollweide projection, which has the polar and azimuthal angles as coordinate axes~\citep{Snyder}. Separately, we also show in Figs.~\ref{fig:4}c and~\ref{fig:4}d both the dipole and quadrupole distributions in the original coordinate system. All the plots are shown at $pH=7$, and a similar figure for HSA is shown in Fig.~\ref{fig:S5} in the Supporting Material.

From the multipole representation of the surface charge distribution where terms up to the order $\ell=6$ are included (Fig.~\ref{fig:4}a) it is not immediately obvious where the dipole and quadrupole axes are located. However, when we isolate both terms (Figs.~\ref{fig:4}c and~\ref{fig:4}d), this becomes more apparent. When the two distributions are combined together with the total charge (Fig.~\ref{fig:4}b), they describe the coarse-grained variation of the surface charge consistent with the Debye screening cut-off. Similar observations can be drawn also for the other proteins studied. For example, by comparing Figs.~\ref{fig:4} and~\ref{fig:S5} we can observe a noticeable difference between $\beta$-lactoglobulin and HSA: In the former, the dipole distribution is dominant among the lower-order multipoles, while the exact opposite is true in the case of the latter.

\begin{figure*}[!b]
\begin{center}
\includegraphics[width=\textwidth]{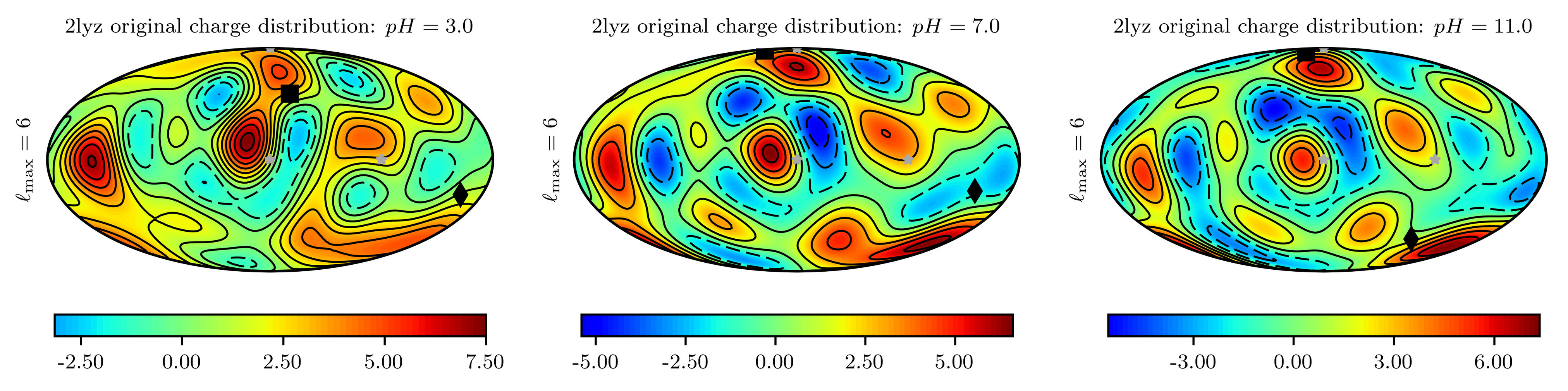}
\end{center}
\caption{Multipole expansion of the surface charge distribution of lysozyme (2lyz) up to $\ell=6$ in the original coordinate system, shown for three different values of $pH=3$, $7$, $11$. The distributions are mapped from a sphere to a plane using the Mollweide projection. Black diamonds show the orientation of the $z$ axis of the dipole, and black squares the orientation of the $z$ axis of the quadrupole. Gray stars show the coordinate axes of the original coordinate system. Cysteine acidity is not considered.
\label{fig:5}}
\end{figure*}

\begin{figure*}[!t]
\begin{center}
\includegraphics[width=0.8\textwidth]{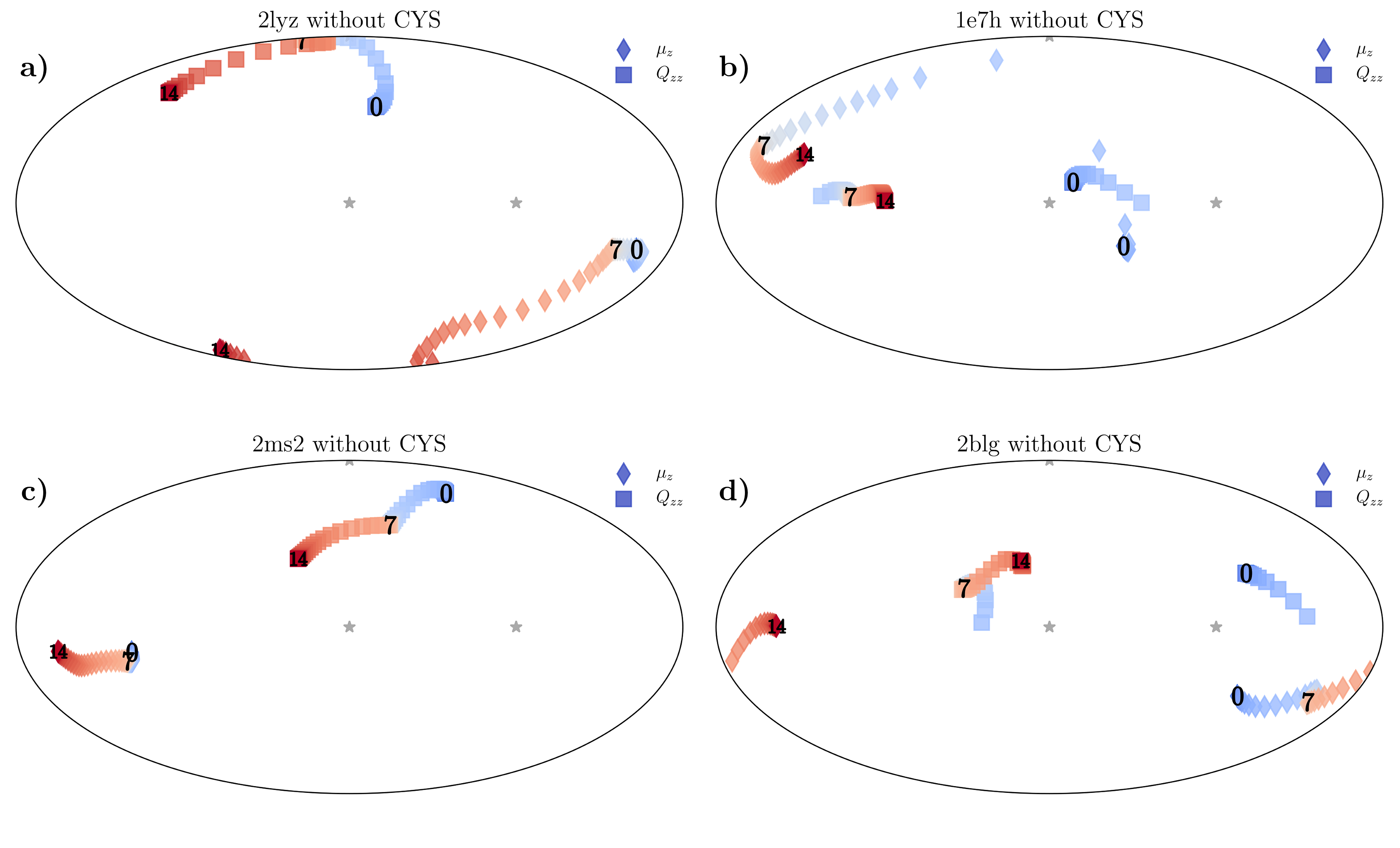}
\end{center}
\caption{{\bf (a)}-{\bf (d)} Orientation of dipole and quadrupole principal $z$ axes (denoted by diamonds and squares, respectively) as a function of $pH$ (in steps of $0.2$ $pH$ unit). The $pH$ increase from 0 to 14 is shown with a color gradient, with blue hues denoting acidic $pH<7$ and red hues denoting basic $pH>7$. In addition, we explicitly indicate the positions at $pH$ values of 0, 7, and 14. The orientations of the axes are mapped from the circumscribed sphere of the protein onto a plane using the Mollweide projection; the gray stars show the coordinate axes of the original coordinate system of the protein, which is kept fixed. The $pH$ dependence of the multipoles' orientation is shown side by side for all four proteins included in our study. Cysteine acidity is not considered.
\label{fig:6}}
\end{figure*}

To demonstrate next how the orientations of the principal $z$ axes of the dipole and quadrupole moments change with $pH$, we show in Fig.~\ref{fig:5} snapshots of the multipole representation of the surface charge distribution of lysozyme (with $\ell_{\max}=6$) at three different values of $pH$. Similar snapshots for HSA are shown in Fig.~\ref{fig:S6} in the Supporting Material. For a complete comparison, Fig.~\ref{fig:6} isolates the orientations of the dipole and quadrupole principal $z$ axes of all four proteins studied, and shows their variation in space over the entire range of $pH$ values while keeping the coordinate system of the protein fixed. (Note that the apparent jumps in the location of the quadrupole axis in the cases of HSA and $\beta$-lactoglobulin are a consequence of our confinement of the axis to the upper hemisphere for reasons of symmetry.)

As $pH$ is increased, the overall charge moves towards more negative values. At the same time, the positions of the dipole and quadrupole principal $z$ axes trace quite a path in space. In the case of the lysozyme (Fig.~\ref{fig:6}a), most of the variation in the axes' orientation occurs after $pH>7$. On the contrary, in the case of HSA (Fig.~\ref{fig:6}b) the majority of the variation, especially large in the case of the dipole, happens up until that point, i.e., when $pH<7$. It is worth noting that the isoelectric points of the two proteins are at the opposite sides of the spectrum (Table~\ref{tab:1}). The axes of the MS2 capsid protein exhibit perhaps the least variation (Fig.~\ref{fig:6}c), while the quadrupole axis of $\beta$-lactoglobulin shifts in space to a large extent (Fig.~\ref{fig:6}d).

Taken together, we see that not only does $pH$ influence the magnitudes of the various multipole moments, but it influences even more so their orientation in space. The changes in the orientation do not happen uniformly with the changing $pH$, but are spaced unevenly across the range. Most of the variation seems to usually (but not always) occur in the $pH$ range near the isoelectric point of each protein.

\section*{DISCUSSION}

The results of our model clearly show that $pH$ variation causes significant changes not only in the magnitudes of charge multipole moments in various proteins, but also in the orientations of the principal axes of their multipoles. In obtaining these results, we have, however, resorted to some necessary simplifications. First of all, while we have used the acid-base dissociation constants $pK_a$ of amino acid functional groups in bulk dilute aqueous solutions, the actual $pK_a$ values in proteins are site-dependent, i.e., influenced by the local ionic and structural environment~\citep{pKvalues,Schaefer1997}. The deviations range from small to significant, depending on the type of site considered~\citep{Gitlin2006,Langlet2008}. Secondly, changes in $pH$ often also lead to conformational changes in proteins~\citep{OBrien2011,Schaefer1997,Harrison2013,Goh2014}, whereas we have used fixed structures obtained from PDB to extract the protein charge distributions. In the following, we discuss these simplifications and limitations of our model, and show that our conclusions should remain valid even as more details are included in the description of charge distributions in proteins.

In our study, we have intentionally de-coupled the effects of conformational changes from the $pH$-dependent changes in charges of a protein, as this allowed us to clearly observe and study the effects of $pH$ on the charge multipoles. In this way we were also able to treat the electrostatics of the proteins using a simple analytical model. Coupling the electrostatics with the conformational changes would, on the other hand, necessarily demand a different treatment -- involving, e.g., studying protein dynamics in order to predict the changes in conformation. At the same time, joining the two models would make it harder to discern which changes in the magnitudes and orientations of the charge multipoles are a consequence of the conformational changes of a protein, and which are a consequence of its $pH$-dependent fractional charges. In the remaining discussion we will thus not delve any further into the influence of conformational changes, but will nonetheless examine the results of our model in light of the effects of local $pK_a$ shifts arising as a consequence of site-site interactions or electrostatic potential. We stress, however, that in real systems where both $pH$-dependent charge and conformational changes operate in parallel, we can expect some quantitative changes in the observed $pH$-dependent behavior of the protein charge multipoles. The discrepancy will naturally be larger in those proteins where conformational changes due to $pH$ are more significant.

To investigate the basic effect of the local $pK_a$ shifts induced by site-site interactions, we used PROPKA software for the empirical prediction of $pK_a$ values of the ionizable amino acids, as described in the Materials and Methods. The results are summarized in Figs.~\ref{fig:S0},~\ref{fig:S0a}, and~\ref{fig:S0b}, and in Table~\ref{tab:S00} in the Supporting Material. Even though the local $pK_a$ values of the amino acid residues now differ from the bulk $pK_a$ values used in our model (Table~\ref{tab:S1}) upward to $2$ units of $pH$, the effect this has on the magnitudes of multipole moments is nonetheless small (Fig.~\ref{fig:S0}). The biggest change can be observed in the case of the phage MS2 capsid protein in the range around neutral $pH$; however, the difference in the predicted isoelectric point is still less than $1$ unit of $pH$ (Table~\ref{tab:S00}). Our model with $pK_a$ values derived from PROPKA also predicts less of a difference between the magnitudes of the dipole and quadrupole moments in the case of $\beta$-lactoglobulin, yet these differences are at most within a factor of $2$ from the predictions of the basic model. Similar differences can be observed with PROPKA-predicted $pK_a$ values in the orientations of the multipole moments, which can range from small in the case of lysozyme (Fig.~\ref{fig:S0a}) to bigger in the case of $\beta$-lactoglobulin (Fig.~\ref{fig:S0b}), yet never significantly change the qualitative behavior of the $pH$ dependence of the multipole orientations.

Another factor that we have not explicitly considered in our model and can nevertheless potentially have a significant effect on the local $pK_a$ values is the electrostatic potential. In principle, the local electrostatic potential at the position of each ionizable amino acids modifies its $pK_a$ value [eq.~\eqref{eq:pn}]. Only when the potential is small is the local $pK_a$ shift due to it negligible. This holds, for instance, in the limit of high salt or large screening, relevant in general for physiological conditions. In order to estimate the $pK_a$ shift due to electrostatic potential in this limit we have extended our model by solving the Debye-H\"{u}ckel equation for the electrostatic potential of the protein in the presence of charge regulation boundary condition. The details of this methodology are given in the Materials and Methods and in the Supporting Material. We observe that the average $pK_a$ shift imparted by the electrostatic potential is on the scale of $1$ $pH$ unit in the limit of high salt concentration ($c_0\sim1$ M), as shown in Fig.~\ref{fig:S00}; this is comparable with the $pK_a$ shifts predicted by PROPKA. In contrast to the latter, the shifts caused by local electrostatic potential are also $pH$ dependent due to the nature of the model. In the limit of high salt, the electrostatically-shifted $pK_a$ values have only a small effect: The predicted isoelectric points are within less than $1$ unit of $pH$ from the predictions of our basic model, and the $pH$-dependence of magnitudes and orientations of the multipole moments follow similar patterns as in the case where electrostatics is not included (Table~\ref{tab:S00} and Figs.~\ref{fig:S0},~\ref{fig:S0a}, and~\ref{fig:S0b}). Notably, the predictions based on the inclusion of the electrostatic potential are sometimes closer to those obtained using PROPKA-predicted $pK_a$ values, while in other cases closer to those obtained using our simple model.

It has to be noted that there are also limitations to the electrostatic model we have used for the prediction of local $pK_a$ shifts. The DH equation performs well in the limit of high salt, which was of interest to us. However, it is known to overpredict the values of the electrostatic potential in the limit of vanishing salt. More importantly, the DH equation is based on a continuum description of the electrostatics. While this is in line with the general model used in our present work, any ion-specific binding effects are thus neglected. Coupling the local description of charge interaction with a global multipole expansion of charge distribution is a difficult problem due to the difference of scales involved. A consistent inclusion of local and ion-specific effects would at this point require a drastically different model, based on an atomic description of the protein and the solvent, while our approach still retains its fundamental validity in the high salt limit.

Although we have not considered detailed structural environments of proteins in our work, we note that structural changes bring about a different configuration of charges on the surface of a protein. A similar re-configuration of charges can arise by the variation of the RSA cut-off $c$. To show that the choice of the RSA cut-off does not have a large impact on the qualitative results of our study, we plot in Fig.~\ref{fig:S7} in the Supporting Material the $pH$ dependence of the multipole moments of the four proteins studied for three different choices of the cut-off, ranging from $c=0.1$ to $c=0.5$. In each of the cases considered, going from $c=0.1$ to $c=0.5$ amounts to quite a significant loss of $24$--$34$\% of the total number of surface charges (see Table~\ref{tab:S1}). And yet, some apparent differences when the cut-off is increased notwithstanding, the qualitative behavior of the multipole moments remains much the same. This validates our approach, which fixed the cut-off to $c=0.25$, and the conclusions drawn from it: While small variations of the number of charges on the surface of a protein will necessarily change the underlying multipole expansion of the surface charge distribution, the qualitative behavior we observed when the $pH$ is varied will remain unaltered.

Lastly, we separately examine the possibility of cysteine protonation, which was not explicitly considered in our main model. Due to its $pK_a$ value, this should in principle push the charge distribution towards more negative values at $pH>7$. In order to see how the consideration of cysteine acidity influences our results, we show in Fig.~\ref{fig:S8} in the Supporting Material a comparison of the $pH$ dependence of the multipole magnitudes in the case of HSA and MS2 capsid protein. We can again observe that the presence of cysteine protonation has no qualitative effect on the behavior of the system. It does affect the location of the isoelectric point to some extent (cf.\ Table~\ref{tab:S00}), although the difference is again less than 1 unit of $pH$. Other than that, the magnitudes of the multipole components and their $pH$ dependence remain relatively unchanged. Similarly, by comparing the orientations of the dipole and quadrupole axes between the cases where cysteine acidity is or is not considered (Figs.~\ref{fig:S9} and~\ref{fig:6}, respectively) we see that the presence of cysteine charges influences the orientational variation of the principal axes at $pH>7$, as expected, although the changes are usually minor.

\section*{CONCLUSIONS}\label{sec:conc}

In this work, we have studied the effects of $pH$ on charge multipole moments in proteins. We first obtained $pH$-dependent surface charge distributions of four globular proteins and expanded the distributions in terms of electrostatic multipoles. We have limited ourselves to lower-order multipoles (the total charge and the dipole and quadrupole moments), and studied the effect of $pH$ on both their magnitudes and the orientations of their principal axes in space. The value of $pH$ was shown to have a significant effect on both, particularly on the orientations of the multipole principal axes, which in some cases exhibited large variation. This variation was found to be a non-uniform function of $pH$, spaced rather unevenly across the solid angle, with most of the changes occurring in the $pH$ range near the isoelectric point of each protein.

We have also pointed out some limitations and necessary simplifications of our approach. While the precise determination of the fractional charge of amino acid residues in a protein due to site-site interactions and local electrostatic potential is difficult, we have nonetheless been able to explore the general influence of local $pK_a$ shifts. We have done this by extending our simple model with either empirically predicted $pK_a$ values or with $pK_a$ values shifted by an electrostatic potential derived from a detailed solution of the corresponding DH equation. In this way, we have shown when and how the qualitative conclusions derived in our paper hold even when the surface charges vary due to local $pK_a$ shifts, and have demonstrated that the observed behavior of multipoles described by our model should persist even as a more detailed picture becomes attainable.

The highly protein-specific $pH$-dependent changes in charge distributions are often intertwined with $pH$-dependent conformational changes of proteins and can occur concomitantly. In order to elucidate the direct influence of the $pH$ dependence of charge multipoles using a simple analytical model, we have neglected any conformational changes; this is, of course, an oversimplification. However, the main results of our model persisted even as the number of charges or their local $pK_a$ values were varied. This indicates that the $pH$ variation of the charge multipole moments of proteins can possibly also play a role as a driving force for conformational changes, and that the $pH$ variation of the direction of the multipole principal axes can affect the electrostatic part of the deformation energy. In this way, it can induce an orientational conformational change of the protein~\citep{OBrien2011,Schaefer1997}, in direct analogy to the structural phase transitions in general condensed matter context.

In addition, the orientational variation of the multipole principal axes remains of fundamental importance for protein-protein interactions also when screened electrostatic interactions in the presence of ionic solution are taken into account. There, the multipole moments play a significantly different role compared to the case of pure, unscreened Couloumb interactions, in the sense that the screened electrostatic potential retains the full directional dependence of all multipole moments even in the far-field limit. Consequently, the usual argument that at large separations only the monopole moment matters is untenable, and while this property of screened multipole electrostatics is not new~\citep{Trizac2000, Kjellander2008, ALB2013a, Kjellander2016}, it has yet to penetrate the community involved in modelling protein-protein interactions.

Finally, we can speculate that higher-order multipole moments can provide a specific electrostatic signature of each protein, and could potentially be used to classify different protein types~\citep{Nakamura1985}. This electrostatic signature could be further refined by defining and tracking the ``orientations'' of higher-order multipoles, possibly allowing for a classification scheme based on a small number of significant parameters coding for the protein symmetry and interactions.

The major lesson of our investigation is thus that $pH$ variation has a significant influence on the magnitudes and especially orientations of charge multipole moments in proteins. This finding was derived using a simple analytical model in a form compatible with models of charged patchy proteins, and at the same time amenable to the inclusion of additional effects, such as empirically predicted $pK_a$ values or $pK_a$ shifts due to local electrostatic potential. The results presented in this work should be of particular significance to protein assembly engineering, e.g., for viral capsids and enzyme nanocontainers, where $pH$ could drive symmetry transitions of the entire assembly induced by local, orientationally-dependent electrostatic interactions.

\section*{SUPPORTING MATERIAL}
Four tables, twelve figures, and thirteen equations of supporting material are available.

\section*{AUTHOR CONTRIBUTIONS}

A.L.B.\ and R.P.\ designed the research; A.L.B.\ performed the research; A.L.B.\ and R.P.\ analyzed the data and wrote the article.

\section*{ACKNOWLEDGMENTS}

We thank S.\ \v{C}opar, G.\ Posnjak, and I.\ Rapo\v{s}ov\'{a} for helpful discussions and comments on the manuscript. We are also grateful to the anonymous reviewers for their comments and suggestions to improve the quality of the manuscript. A.L.B.\ and R.P.\ acknowledge the financial support from the Slovenian Research Agency (research core funding No.\ (P1-0055)).

\bibliographystyle{biophysj}
\bibliography{references}

\newpage
\setcounter{figure}{0}
\setcounter{table}{0}
\setcounter{equation}{0}
\renewcommand{\thefigure}{S\arabic{figure}}
\renewcommand{\theequation}{S\arabic{equation}}
\renewcommand{\thetable}{S\arabic{table}}

\vspace{0.5cm}
\centerline{\large{{\bf Supporting Material: pH dependence of charge multipole moments in proteins}}}
\vspace{0.5cm}
\centerline{\normalsize{A.\ Lo\v{s}dorfer Bo\v{z}i\v{c} and R.\ Podgornik}}

\newpage

\section{NUMBER OF SURFACE CHARGES AS A FUNCTION OF RSA CUT-OFF}

\begin{table}[!htp]
\begin{center}
\begin{tabular}{lcccccc}
 & 0.1 & 0.2 & {\bf 0.25} & 0.3 & 0.4 & 0.5 \\
\hline
\hline
2lyz (no CYS) & 30 & 30 & {\bf 30} & 28 & 26 & 23 \\
2lyz (with CYS) & 34 & 33 & {\bf 33} & 31 & 27 & 24 \\
1e7h (no CYS) & 164 & 162 & {\bf 157} & 152 & 145 & 119 \\
1e7h (with CYS) & 192 & 188 & {\bf 182} & 173 & 162 & 127 \\
2ms2 (no CYS) & 23 & 23 & {\bf 22} & 22 & 19 & 17 \\
2ms2 (with CYS) & 25 & 25 & {\bf 24} & 24 & 21 & 19 \\
2blg (no CYS) & 50 & 49 & {\bf 46} & 45 & 42 & 36 \\
2blg (with CYS) & 51 & 50 & {\bf 47} & 46 & 43 & 37
\end{tabular}
\end{center}
\caption{Number of charges present on the solvent-exposed surfaces of proteins used in our study as a function of the RSA cut-off $c$, with or without considering the cysteine (CYS) acidity. The number of charges present on the surface decreases with increasing cut-off. Outlined in bold is the cut-off of $c=0.25$, used in the majority of the main text.
\label{tab:S1}}
\end{table}

\section{ACID-BASE EQUILIBRIUM CONSTANTS}

\begin{table}[!htp]
\begin{center}
\begin{tabular}{cccccccc}
 & ASP & GLU & TYR & ARG & HIS & LYS & CYS \\
\hline
\hline
$pK_a$ & 3.71 & 4.15 & 10.10 & 12.10 & 6.04 & 10.67 & 8.14
\end{tabular}
\end{center}
\caption{Intrinsic $pK_a$ values of amino acid functional groups in bulk dilute aqueous solutions. Values taken from Ref.~\citep{CRC}.
\label{tab:S2}}
\end{table}

\section{CIRCUMSCRIBED RADII OF STUDIED PROTEINS}

\begin{table}[!htp]
\begin{center}
\begin{tabular}{lcccc}
 & 2lyz & 1e7h & 2blg & 2ms2 \\
\hline
\hline
$R$ [nm] & 0.987 & 1.97 & 1.06 & 1.12
\end{tabular}
\end{center}
\caption{Circumscribed radii of the proteins used in the study: lysozyme (2lys), human serum albumin (1e7h), $\beta$-lactoglobulin (2blg), and subunit A of the phage MS2 capsid protein (2ms2). The radii were obtained as the largest distance of any $C_\alpha$ atom from the center-of-mass of a protein.
\label{tab:S3}}
\end{table}

\section{ROLE OF ELECTROSTATIC POTENTIAL IN CHARGE REGULATION}

Local electrostatic potential $\psi(\mathbf{r})$ can play a role in the regulation of charge on the amino acid residues, inducing a local $pK_a$ shift. The effect of the potential on the charge of an amino acid can be written as~\cite{Ninham1971}
\begin{equation}
\label{eq:chr}
q^{\pm}_k=\frac{\pm e_0}{1+e^{\pm\ln10(pH-pK_a^{(k)})\mp\beta e_0\psi(\mathbf{r}_k)}},
\end{equation}
where $\beta=1/k_BT$, $T$ is the room temperature, $k_B$ the Boltzmann constant, and $e_0$ the elementary charge. The renormalized $pK_a$ values thus become
\begin{equation}
\label{eq:shift}
pK_a^{(k)}\rightarrow pK_a^{(k)}+\frac{\beta e_0\psi(\mathbf{r}_k)}{\ln10}.
\end{equation}

In order to obtain the electrostatic potential, we use a reformulation of the canonical Tanford-Kirkwood model~\citep{TanfordKirkwood, Fer2001} by treating the protein as a sphere with radius $R$ and dielectric constant $\varepsilon_p=4$, carrying a surface charge distribution $\sigma(\Omega)$ including the monopole, dipole, and quadrupole contributions ($\ell_\mathrm{max}=2$), and immersed in a $1:1$ salt solution with bulk concentration $c_0$ and with the dielectric constant of water $\varepsilon_w=80$. At large, physiologically relevant salt concentrations ($c_0\gtrsim100$ mM), we can solve the Poisson-Boltzmann equation for the electrostatic potential in the linearized, Debye-H\"{u}ckel (DH) approximation:
\begin{equation}
\label{eq:dh}
\nabla^2\psi_1(r,\Omega)=0\quad ;\;r\leq R\quad\mathrm{and}\quad\nabla^2\psi_2(r,\Omega)=\kappa^2\psi_2(r,\Omega)\quad ;\;r\geq R,
\end{equation}
inside and outside of the protein, respectively (for details, see for instance Ref.~\cite{ALB2013a}). Here, $\kappa=\sqrt{8\pi\ell_B c_0}$ is the inverse Debye length, and $\ell_B=\beta e_0^2/4\pi\varepsilon_0\varepsilon_w$ is the Bjerrum length. The two potentials are continuous on the boundary of the protein,
\begin{equation}
\psi_1(R,\Omega)=\psi_2(R,\Omega)\quad\forall\Omega,
\end{equation}
while the jump in their derivatives is proportional to the charge on the protein, which is in turn coupled with the local potential
\begin{equation}
\label{eq:surf}
\varepsilon_p\varepsilon_0\frac{\partial\psi_1(r,\Omega)}{\partial r}\Big|_{r=R}-\varepsilon_w\varepsilon_0\frac{\partial\psi_2(r,\Omega)}{\partial r}\Big|_{r=R}=\sigma(\Omega)=\sigma^{(0)}(\Omega)+\sigma^{(1)}(\Omega)\beta e_0\psi(R,\Omega)\quad\forall\Omega.
\end{equation}
Here, we have used the high-salt/low electrostatic potential approximation to expand the regulated charge in eq.~\eqref{eq:chr} to the linear order in $\beta e_0\psi\ll1$:
\begin{equation}
q_k^\pm\approx\frac{\pm 1}{1+e^{\pm\ln10\,\left(pH-pK_a^{(k)}\right)}}\mp\frac{e^{\pm\ln10\,\left(pH-pK_a^{(k)}\right)}}{\left[1+e^{\pm\ln10\,\left(pH-pK_a^{(k)}\right)}\right]^2}\times\beta e_0\psi(R,\Omega_k)=q_k^{\pm,(0)}+q_k^{\pm,(1)}\beta e_0\psi(R,\Omega_k).
\end{equation}
This in turn yields a surface charge expansion of the form
\begin{equation}
\sigma(\Omega)=\sigma^{(0)}(\Omega)+\sigma^{(1)}(\Omega)\beta e_0\psi(R,\Omega)=\frac{e_0}{R^2}\left[\sum_{k\in\mathrm{AA}}q_k^{(0)}\delta(\Omega-\Omega_k)+\sum_{k\in\mathrm{AA}}q_k^{(1)}\delta(\Omega-\Omega_k)\beta e_0\psi(R,\Omega)\right],
\end{equation}
where both terms can be expanded in terms of multipoles,
\begin{equation}
\sigma_{lm}^{(0)}=\sum_{k\in\mathrm{AA}}q_k^{(0)}Y_{lm}^*(\Omega_k)\quad\mathrm{and}\quad\sigma_{lm}^{(1)}=\sum_{k\in\mathrm{AA}}q_k^{(1)}Y_{lm}^*(\Omega_k).
\end{equation}
We note that the first term of the expanded surface charge density, $\sigma^{(0)}(\Omega)$, corresponds to the original surface charge density used in the main text which is not corrected for the effects of electrostatic charge regulation.

The DH equation for the electrostatic potential inside and outside the protein [eq.~\eqref{eq:dh}] can be solved in the form
\begin{equation}
\psi_1=\sum_{l,m} A_{lm}\,\left(\frac{r}{R}\right)^l\,Y_{lm}(\Omega)\quad\mathrm{and}\quad\psi_2=\sum_{l,m} B_{lm}\,k_l(\kappa r)\,Y_{lm}(\Omega),
\end{equation}
where $k_l(x)$ are the modified spherical Bessel functions of the second kind. By equating the two potentials at $r=R$ and taking into account the boundary condition for the derivative of the potential at $r=R$, we obtain the following expression for the coefficients $B_{uv}$:
\begin{equation}
\label{eq:blm}
f_u(\kappa R)B_{uv}=\sigma_{uv}^{(0)}+\beta e_0\sum_{l,m}\sum_{l',m'}\sigma_{lm}^{(1)}\,B_{l'm'}\,k_{l'}(\kappa R)\,T(uv|lm|l'm').
\end{equation}
Here, we have introduced a shorthand
\begin{equation}
f_l(\kappa R)=\frac{R^2\kappa\varepsilon_w\varepsilon_0}{e_0}\left[\frac{\varepsilon_p}{\varepsilon_w}\frac{l\,k_l(\kappa R)}{\kappa R}-k'_l(\kappa R)\right],
\end{equation}
and used the standard form of the Wigner 3-j symbols (for details, see again Ref.~\cite{ALB2013a})
\begin{equation}
\label{eq:fin}
T(uv|lm|l'm')=\sqrt{\frac{(2u+1)(2l+1)(2l'+1)}{4\pi}}\times\tj{u}{l}{l'}{0}{0}{0}\tj{u}{l}{l'}{v}{m}{m'}.
\end{equation}
Due to charge regulation on the surface of the protein, the different coefficients $B_{uv}$ are clearly coupled, and eq.~\eqref{eq:blm} cannot be solved further analytically. However, an iterative scheme with the initial value $B_{uv}^0=\sigma_{uv}^{(0)}/f_u(\kappa R)$ allows us to obtain the coefficients $B_{uv}$ with a desired accuracy in typically 5-10 iterations.

\begin{figure*}[!htp]
\begin{center}
\includegraphics[width=0.48\textwidth]{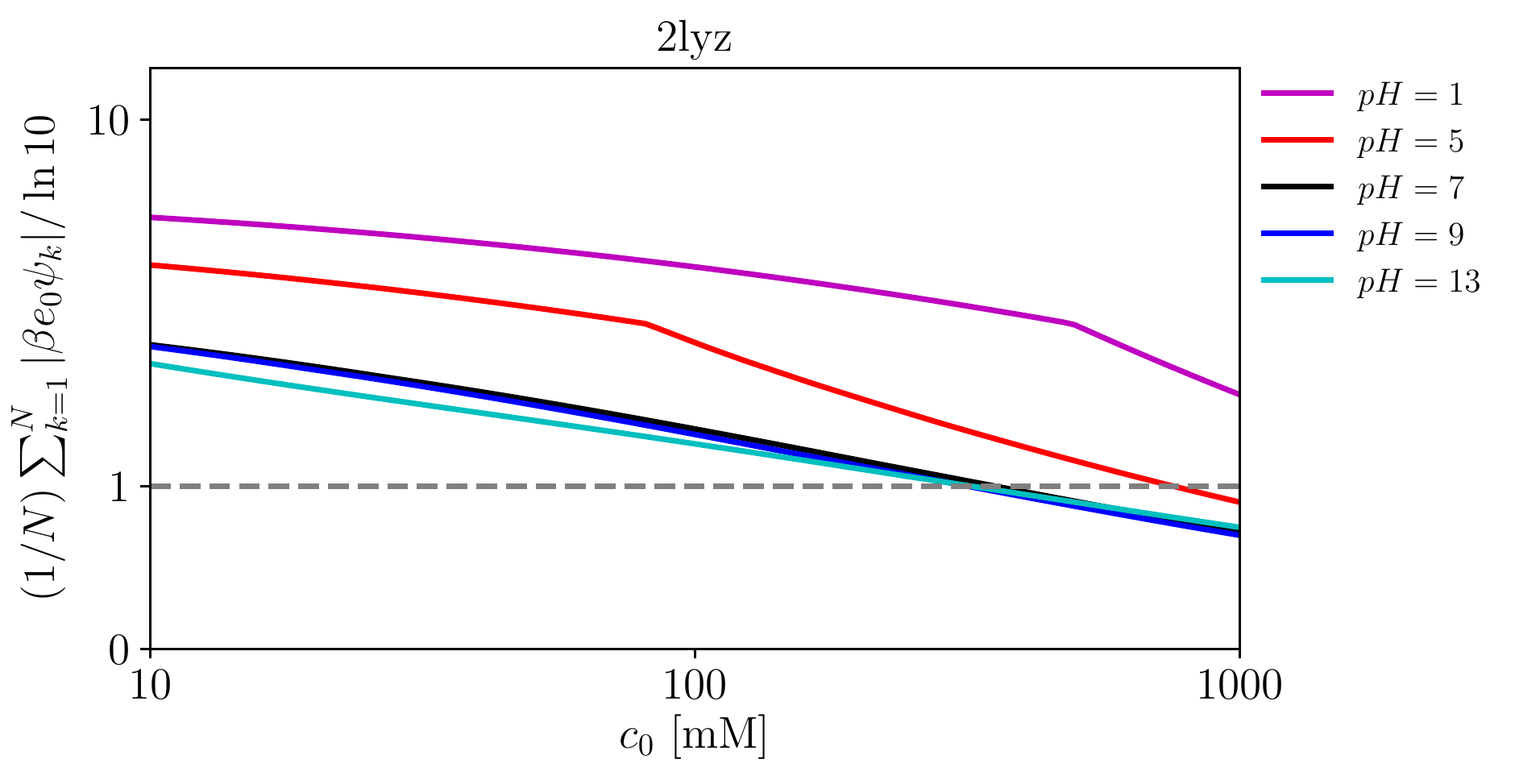}
\includegraphics[width=0.48\textwidth]{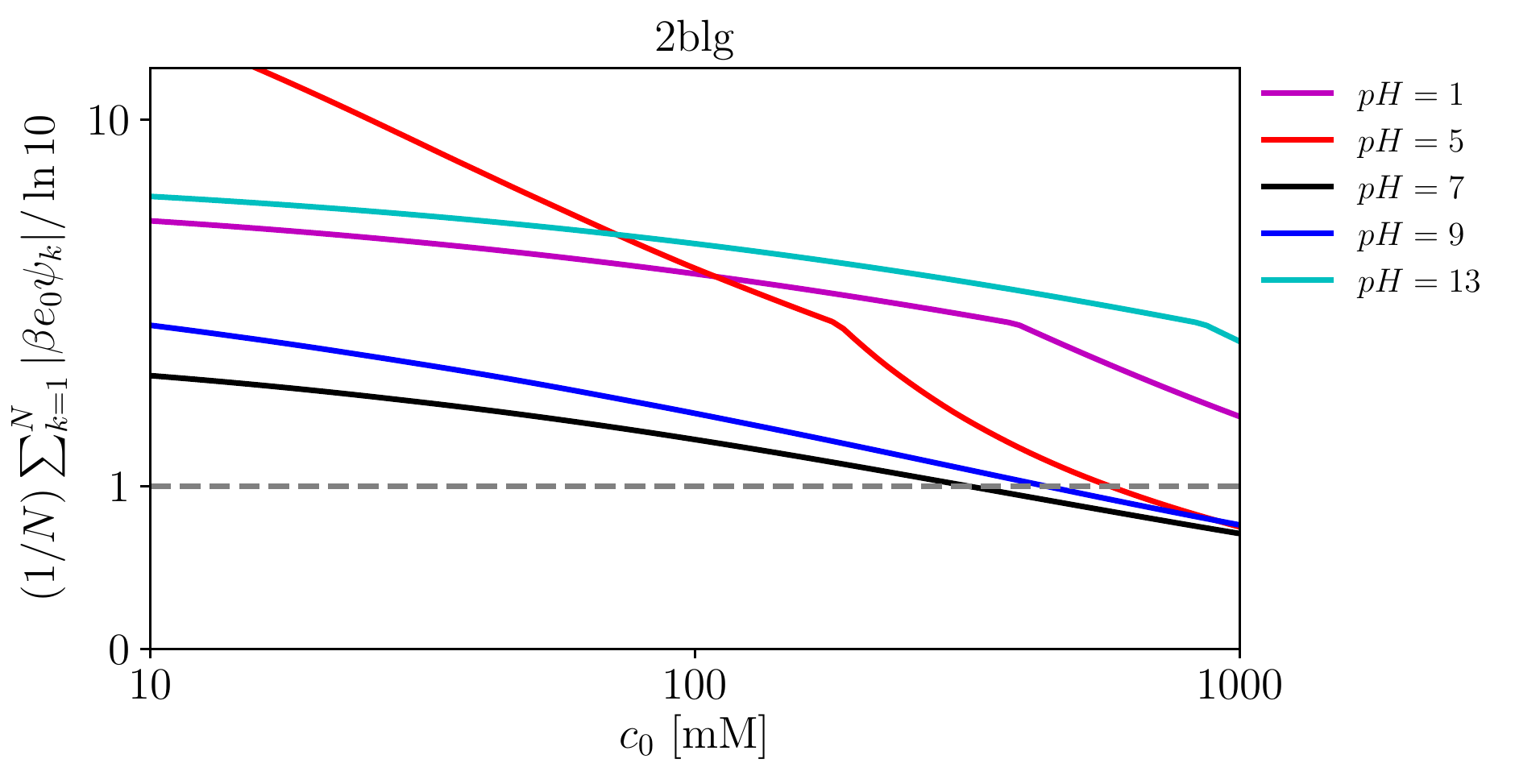}
\end{center}
\caption{Mean $pK_a$ shift due to the DH electrostatic potential at the position of each amino acid, averaged over all $N$ ionizable amino acids in the protein, shown as a function of the bulk salt concentration, $c_0$. As the electrostatic potential and thus the $pK_a$ shift are $pH$-dependent, we plot the mean $pK_a$ shift for 5 different values of $pH$. Shown for lysozyme (2lyz) and $\beta$-lactoglobulin (2blg).
\label{fig:S00}}
\end{figure*}

The solution for the coefficients $B_{lm}$ allows us to obtain the local electrostatic potential on the surface of the protein in the continuum approximation. Evaluating the potential at the positions of the amino acid residues enables us to write their re-normalized charges as given by eq.~\eqref{eq:chr} and to obtain the $pK_a$ shifts of individual amino acids [eq.~\eqref{eq:shift}]. We also note that the $pK_a$ shifts obtained in this way are $pH$-dependent due to the way the electrostatic potential is coupled with the fractional charge of the amino acids. In order to show the average effect the electrostatic potential has on the local $pK_a$ values, we plot in Fig.~\ref{fig:S00} the mean $pK_a$ shift, averaged over all $N$ ionizable amino acids in a protein,
\begin{equation}
\overline{pK}_a=\frac{1}{N}\sum_{k\in\mathrm{AA}}\frac{|\beta e_0\psi(\mathbf{r}_k)|}{\ln10}.
\end{equation}
 
\clearpage
\newpage

\section{GEOMETRICAL INTERPRETATION OF THE QUADRUPOLE EIGENVALUES RATIO}

\begin{figure*}[!htp]
\begin{center}
\includegraphics[width=\textwidth]{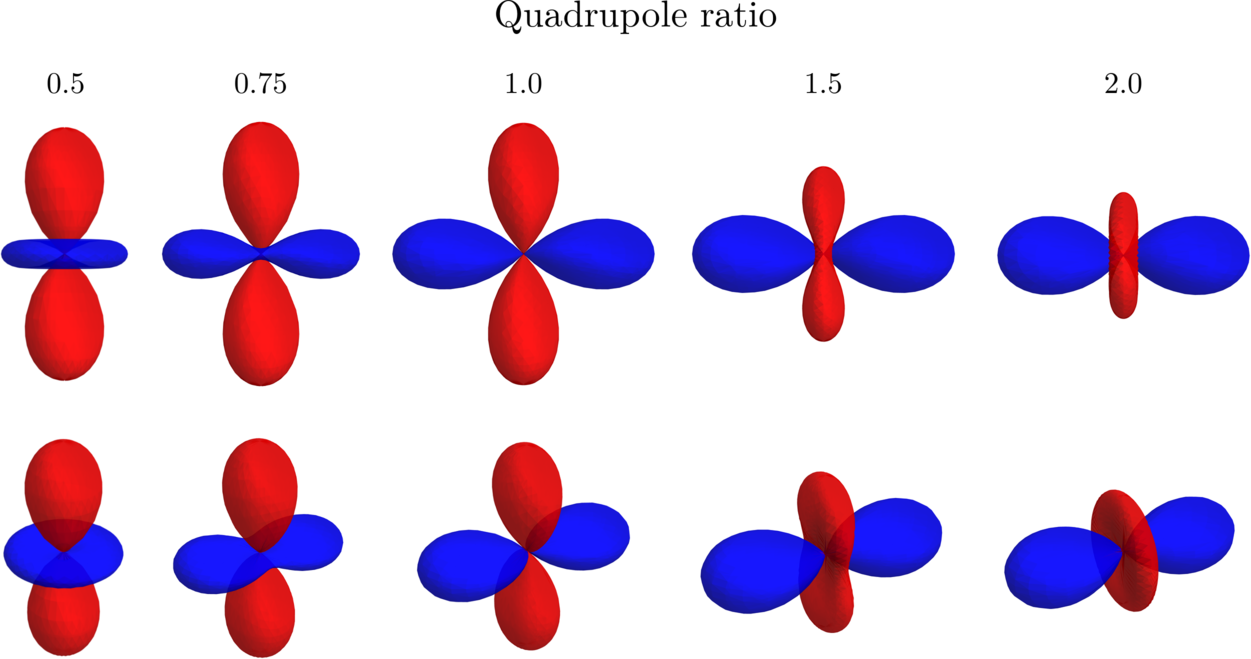}
\end{center}
\caption{Illustration of quadrupole moment distributions for different values of the quadrupole ratio $|Q_{xx}/Q_{zz}|$ ($0.5$-$2$, left to right). The top and bottom rows show two different spatial perspectives of each ratio. Red color denotes positive values, blue negative, and the radius of the distribution corresponds to its magnitude. In the first of the extreme cases we have $|Q_{xx}/Q_{zz}|=0.5$ and thus $Q_{xx}=Q_{yy}$. This corresponds to the case of an axial molecule, where the quadrupole is oriented along the $z$ axis with a symmetric belt of opposite charge located in the $x$-$y$ plane. A similar case is observed when $|Q_{xx}/Q_{zz}|=2$ and the distribution is oriented along the $x$ axis with $Q_{zz}=Q_{yy}$. In the other extreme, we have $|Q_{xx}/Q_{zz}|=1$ and thus $Q_{xx}=-Q_{zz}$ with $Q_{yy}=0$. In this case, the quadrupole distribution is symmetric in the $x$-$z$ plane and vanishes completely along the $y$ axis. See also Fig.~\ref{fig:S2} for some of these distributions projected onto a plane.
\label{fig:S1}}
\end{figure*}

\begin{figure*}[!htp]
\begin{center}
\includegraphics[width=0.32\textwidth]{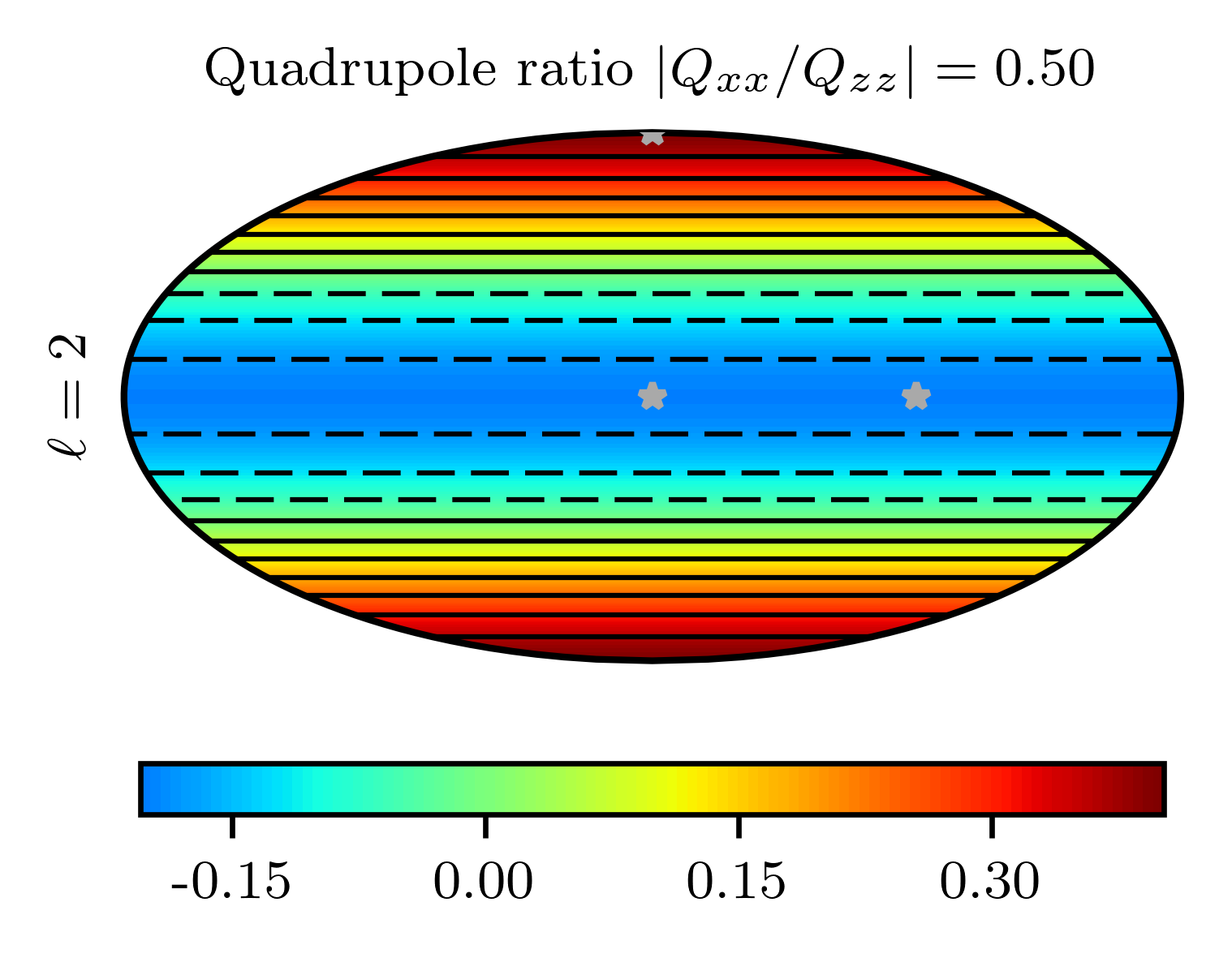}
\includegraphics[width=0.32\textwidth]{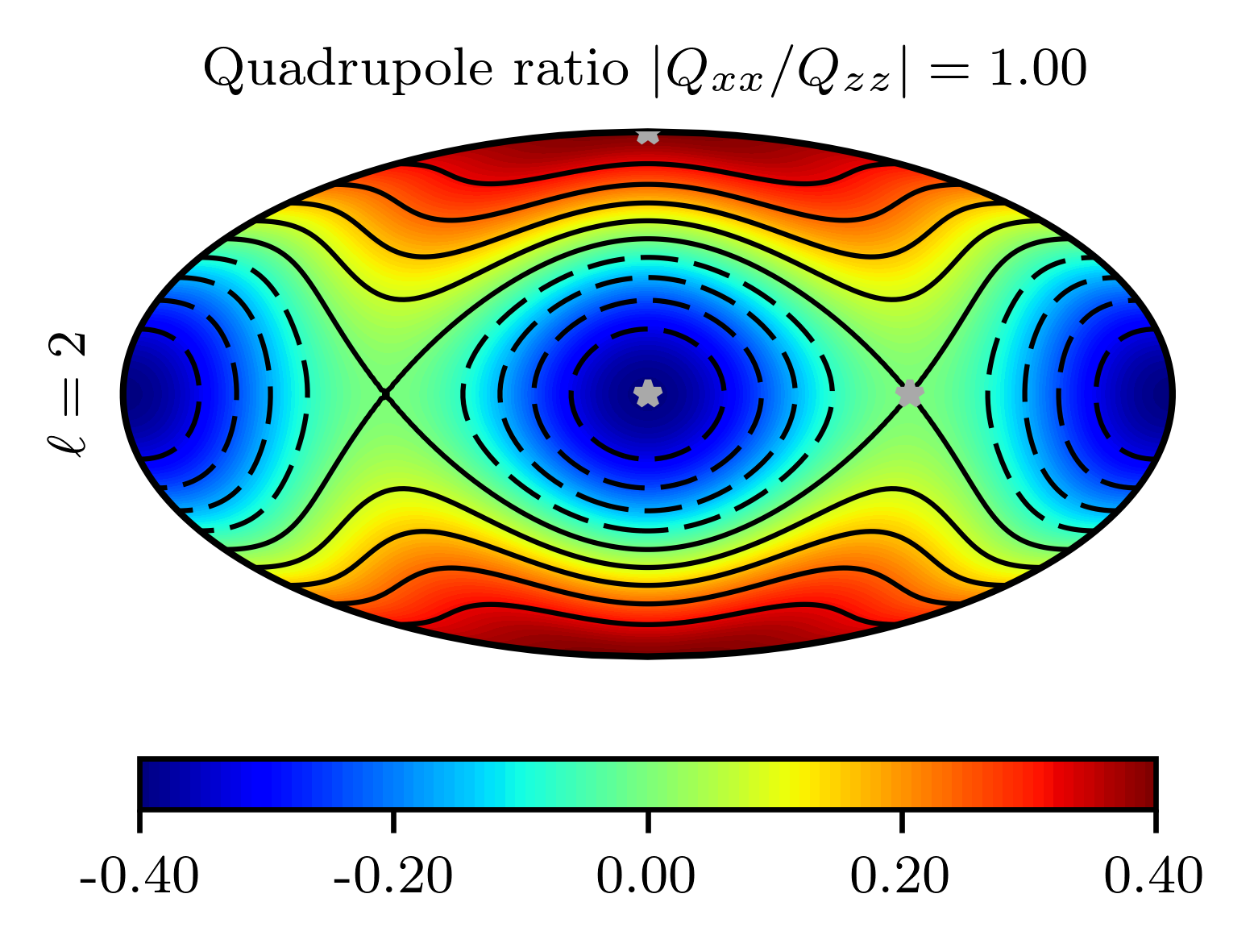}
\includegraphics[width=0.32\textwidth]{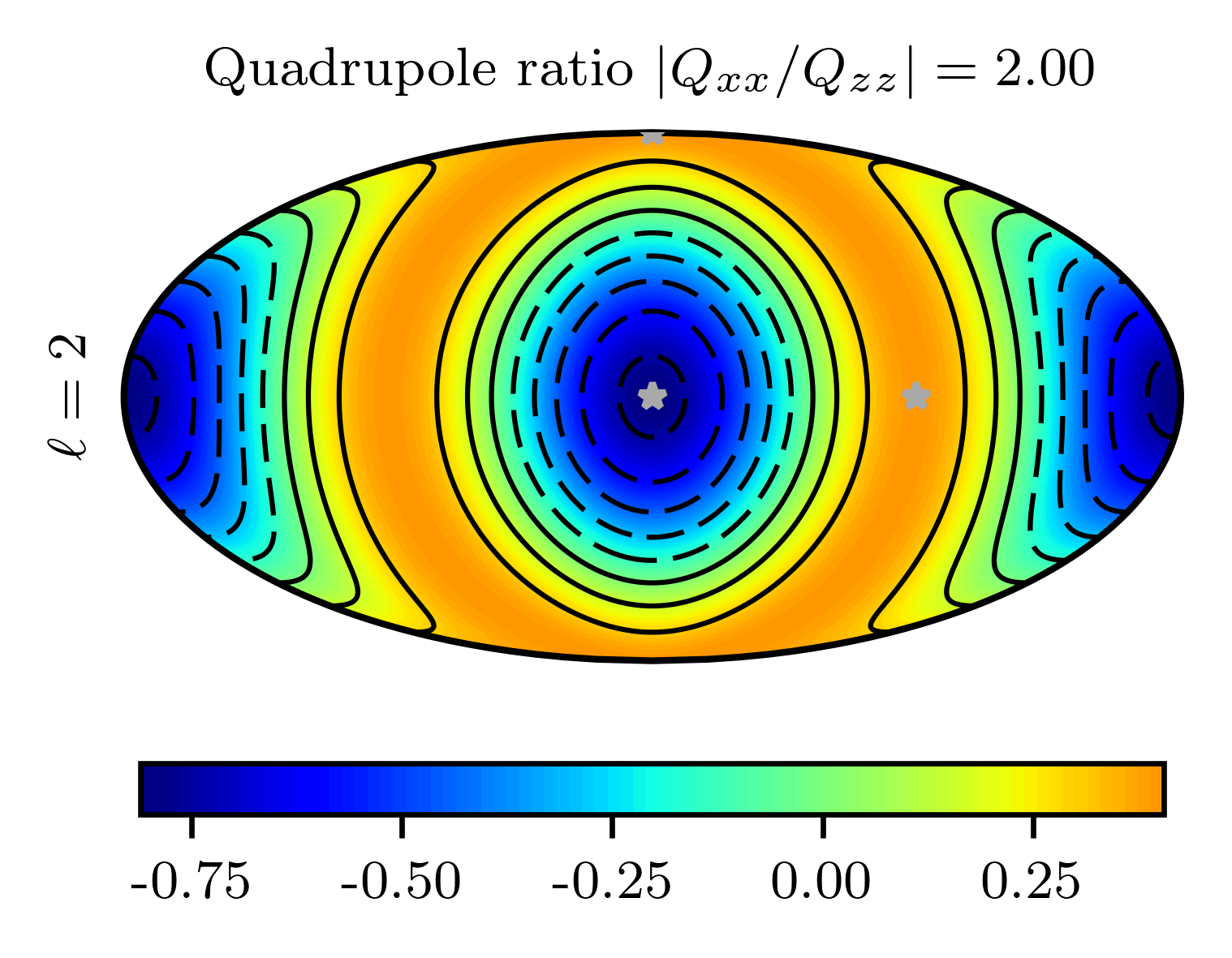}
\end{center}
\caption{Quadrupole distributions with $Q_{zz}=1$ and different values of the quadrupole ratio $|Q_{xx}/Q_{zz}|$. The distributions are mapped from a sphere to a plane using the Mollweide projection. The axes of the coordinate system are shown in gray. See Fig.~\ref{fig:S1} for details.
\label{fig:S2}}
\end{figure*}

\clearpage
\newpage

\section{PH DEPENDENCE OF MULTIPOLE MAGNITUDES}

\begin{figure*}[!htp]
\begin{center}
\includegraphics[width=0.48\textwidth]{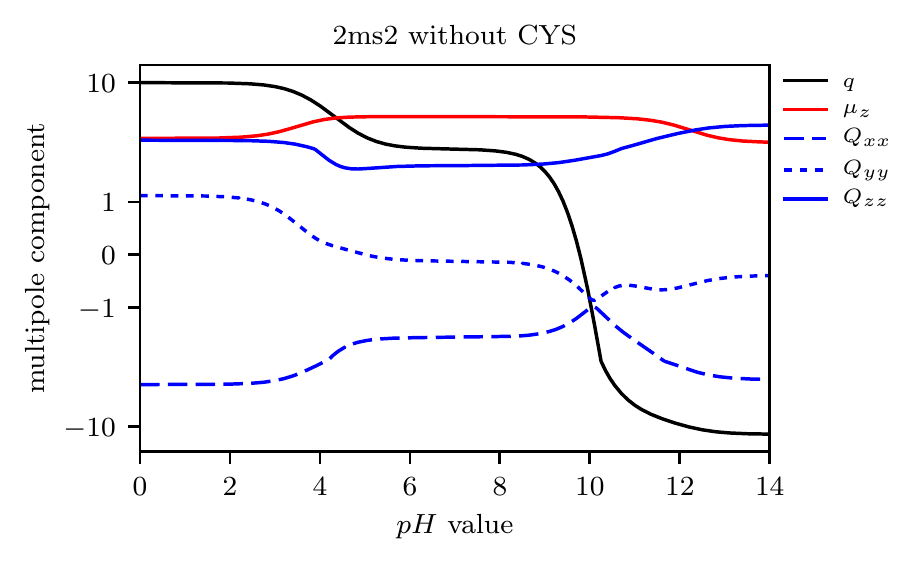}
\includegraphics[width=0.48\textwidth]{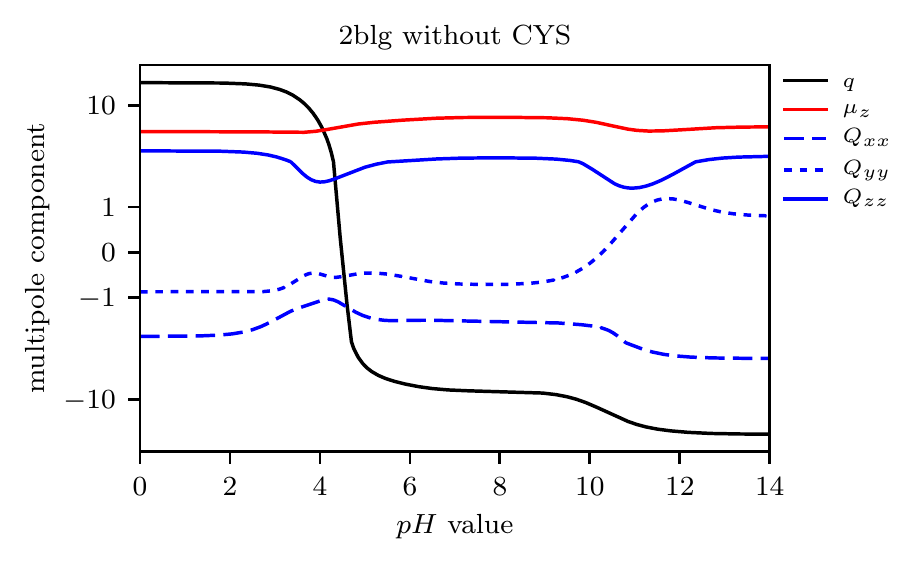}
\end{center}
\caption{Magnitudes of the monopole ($q$), dipole ($\mu_z$), and quadrupole ($Q_{ii}$) components of the surface charge distributions of the MS2 capsid protein (2ms2) and $\beta$-lactoglobulin (2blg), shown as a function of $pH$. Cysteine acidity is not considered.
\label{fig:S4}}
\end{figure*}

\clearpage
\newpage

\section{PH DEPENDENCE OF DIPOLE AND QUADRUPOLE PRINCIPAL AXES}

\begin{figure*}[!htp]
\begin{center}
\includegraphics[width=0.8\textwidth]{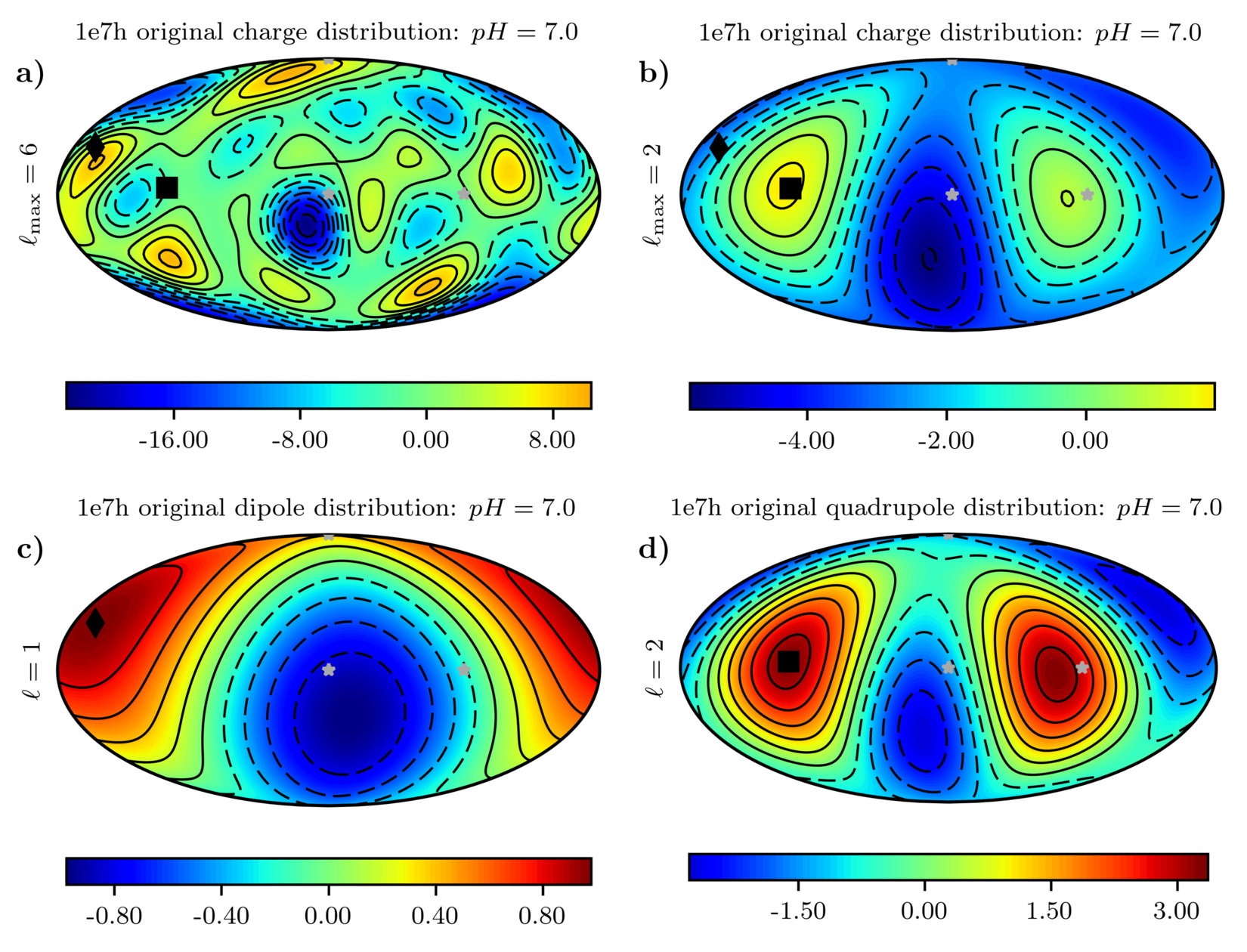}
\end{center}
\caption{Multipole expansion of the surface charge distribution of human serum albumin (1e7h) in the original coordinate system up to {\bf (a)} $\ell_{\max}=6$ and {\bf (b)} $\ell_{\max}=2$. Shown are also the {\bf (c)} dipole and {\bf (d)} quadrupole distributions in the original (reference) system. The distributions are mapped from a sphere to a plane using the Mollweide projection. Black diamonds show the orientation of the $z$ axis of the dipole, and black squares the orientation of the $z$ axis of the quadrupole. Gray stars show the coordinate axes of the original coordinate system. Cysteine acidity is not considered, and all the plots are drawn at $pH=7.0$.
\label{fig:S5}}
\end{figure*}

\begin{figure*}[!htp]
\begin{center}
\includegraphics[width=0.325\textwidth]{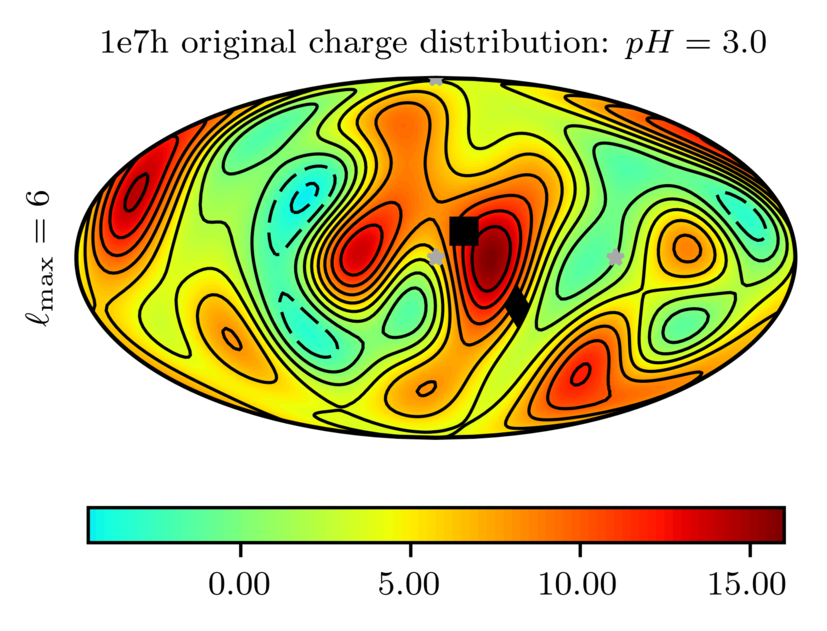}
\includegraphics[width=0.325\textwidth]{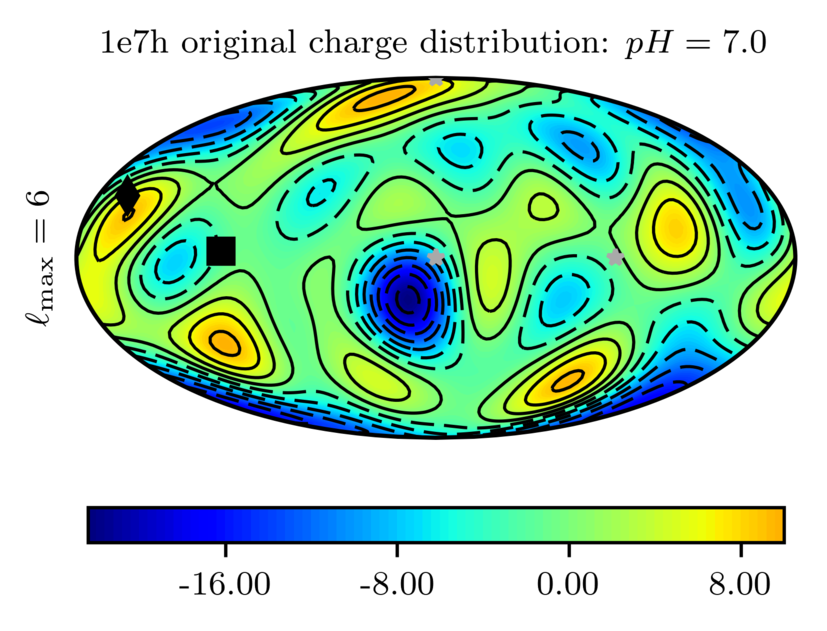}
\includegraphics[width=0.325\textwidth]{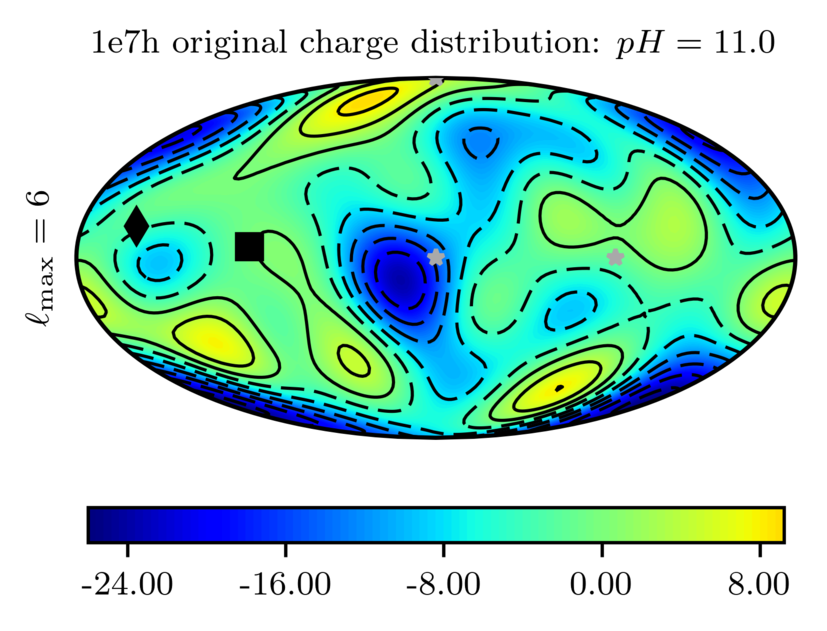}
\end{center}
\caption{Multipole expansion of the surface charge distribution of human serum albumin (1e7h) up to $\ell=6$ in the original coordinate system, shown for three different values of $pH=3$, $7$, $11$. The distributions are mapped from a sphere to a plane using the Mollweide projection. Black diamonds show the orientation of the $z$ axis of the dipole, and black squares the orientation of the $z$ axis of the quadrupole. Gray stars show the coordinate axes of the original coordinate system. Cysteine acidity is not considered.
\label{fig:S6}}
\end{figure*}

\clearpage
\newpage

\section{COMPARISON WITH PROPKA AND ELECTROSTATIC CHARGE REGULATION}

\begin{table*}[!htp]
\begin{center}
\begin{tabular}{lcccccccc}
PDB & model & $pI$ & $\mu_I$ $[e_0\times\textrm{nm}]$ & $Q_I$ $[e_0\times\textrm{nm}^2]$ & $q_n$ $[e_0]$ & $\mu_n$ $[e_0\times\textrm{nm}]$ & $Q_n$ $[e_0\times\textrm{nm}^2]$ & $\langle|Q_{xx}/Q_{zz}|\rangle_{pH}$ \\
\hline
\hline
2lyz & no CYS & $11.08$ & $1.85$ & $3.86$ & $8.10$ & $2.67$ & $2.92$ & $1.09$ \\
2lyz & CYS & $10.53$ & $1.80$ & $3.18$ & $7.90$ & $2.61$ & $2.82$ & $1.24$ \\
2lyz & PROPKA & $11.38$ & $2.60$ & $3.85$ & $8.70$ & $4.43$ & $3.75$ & $1.16$ \\
2lyz & $c_0=1$ M & $11.08$ & $2.60$ & $5.16$ & $8.38$ & $2.75$ & $3.08$ & $1.08$ \\
\hline
1e7h & no CYS & $4.78$ & $2.27$ & $34.68$ & $-23.45$ & $7.86$ & $38.01$ & $0.93$ \\
1e7h & CYS & $4.78$ & $2.27$ & $34.69$ & $-25.14$ & $8.08$ & $39.17$ & $0.95$ \\
1e7h & PROPKA & $5.12$ & $4.40$ & $31.23$ & $-19.27$ & $8.61$ & $33.14$ & $0.88$ \\
1e7h & $c_0=1$ M & $4.78$ & $3.91$ & $53.41$ & $-24.41$ & $8.61$ & $39.47$ & $0.93$ \\
\hline
2blg & no CYS & $4.47$ & $5.77$ & $2.01$ & $-7.78$ & $7.43$ & $2.64$ & $0.91$ \\
2blg & CYS & $4.47$ & $5.77$ & $2.01$ & $-7.85$ & $7.49$ & $2.58$ & $0.97$ \\
2blg & PROPKA & $4.78$ & $3.95$ & $4.28$ & $-8.62$ & $5.41$ & $5.77$ & $1.19$ \\
2blg & $c_0=1$ M & $4.38$ & $10.46$ & $4.26$ & $-8.00$ & $7.55$ & $2.70$ & $0.92$ \\
\hline
2ms2 & no CYS & $9.78$ & $4.88$ & $2.34$ & $2.00$ & $4.91$ & $2.78$ & $1.02$ \\
2ms2 & CYS & $8.88$ & $4.65$ & $2.64$ & $1.87$ & $4.87$ & $2.68$ & $1.07$ \\
2ms2 & PROPKA & $9.03$ & $4.22$ & $1.49$ & $1.00$ & $4.35$ & $2.19$ & $1.24$ \\
2ms2 & $c_0=1$ M & $9.68$ & $5.55$ & $2.47$ & $2.01$ & $4.92$ & $2.78$ & $1.03$
\end{tabular}
\end{center}
\caption{Summary of results for the $pH$ dependence of the magnitudes of multipole components, comparing our simple model, where every amino acid group is assigned the same $pK_a$ value [cf.\ Table~\ref{tab:S2}], with two models introducing local effects to the $pK_a$ values of individual amino acid residues. From the latter two models, the first one uses empirical predictions of PROPKA software, which includes site-site interactions~\cite{Sondergaard2011}. The second model takes into account the local $pK_a$ shifts imparted by the electrostatic potential of the protein in the presence of 1:1 salt with bulk concentration $c_0=1$ M, derived using linearized DH equation with charge regulation boundary condition. Listed are the isoelectric point, $pI$, where the monopole moment (total charge) vanishes, and the magnitudes of the dipole and quadrupole moment in this point (subscript $I$). We also list the magnitudes of the monopole, dipole, and quadrupole moments at neutral $pH=7$ (subscript $n$). Lastly, we list the quadrupole ratio $\langle|Q_{xx}/Q_{zz}|\rangle_{pH}$, averaged across the entire range of $pH$ values. The magnitudes of the multipole moments correspond to total charge in the case of the monopole, Cartesian norm in the case of the dipole, and $Q=[(Q_{xx}-Q_{yy})^2+Q_{zz}^2]^{1/2}$ in the case of the quadrupole.
\label{tab:S00}}
\end{table*}

\begin{figure*}[!htp]
\begin{center}
\includegraphics[width=0.48\textwidth]{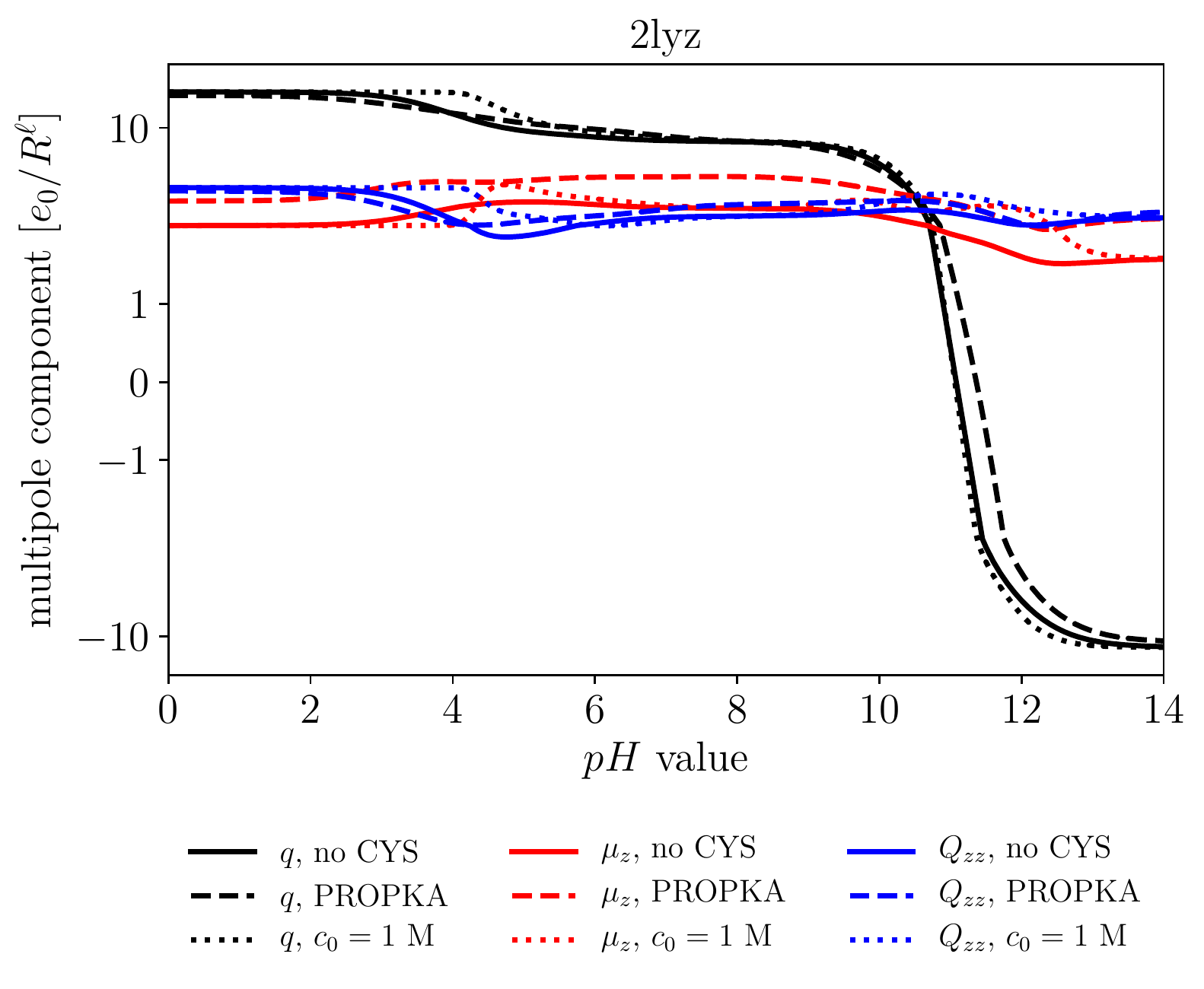}
\includegraphics[width=0.48\textwidth]{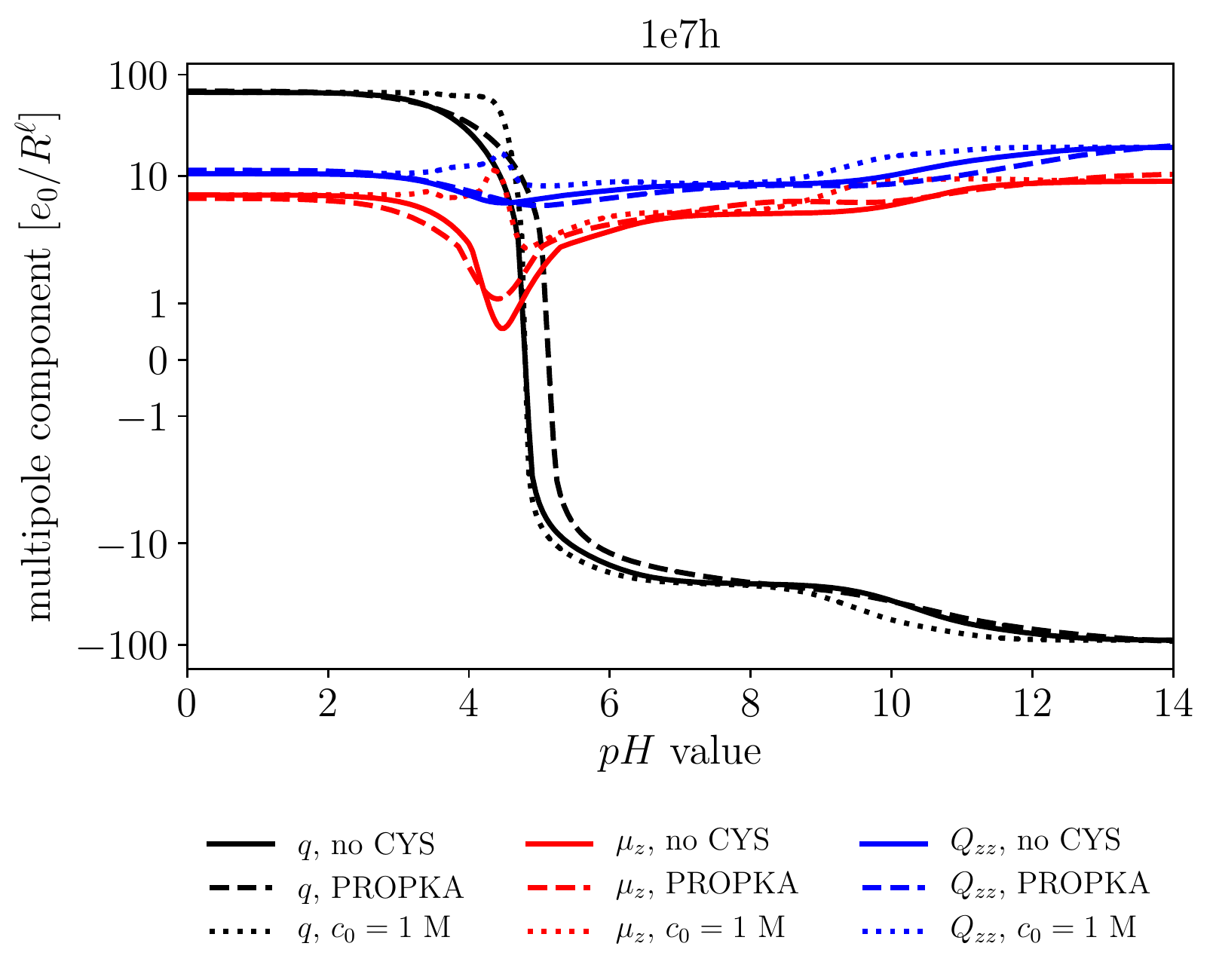}
\includegraphics[width=0.48\textwidth]{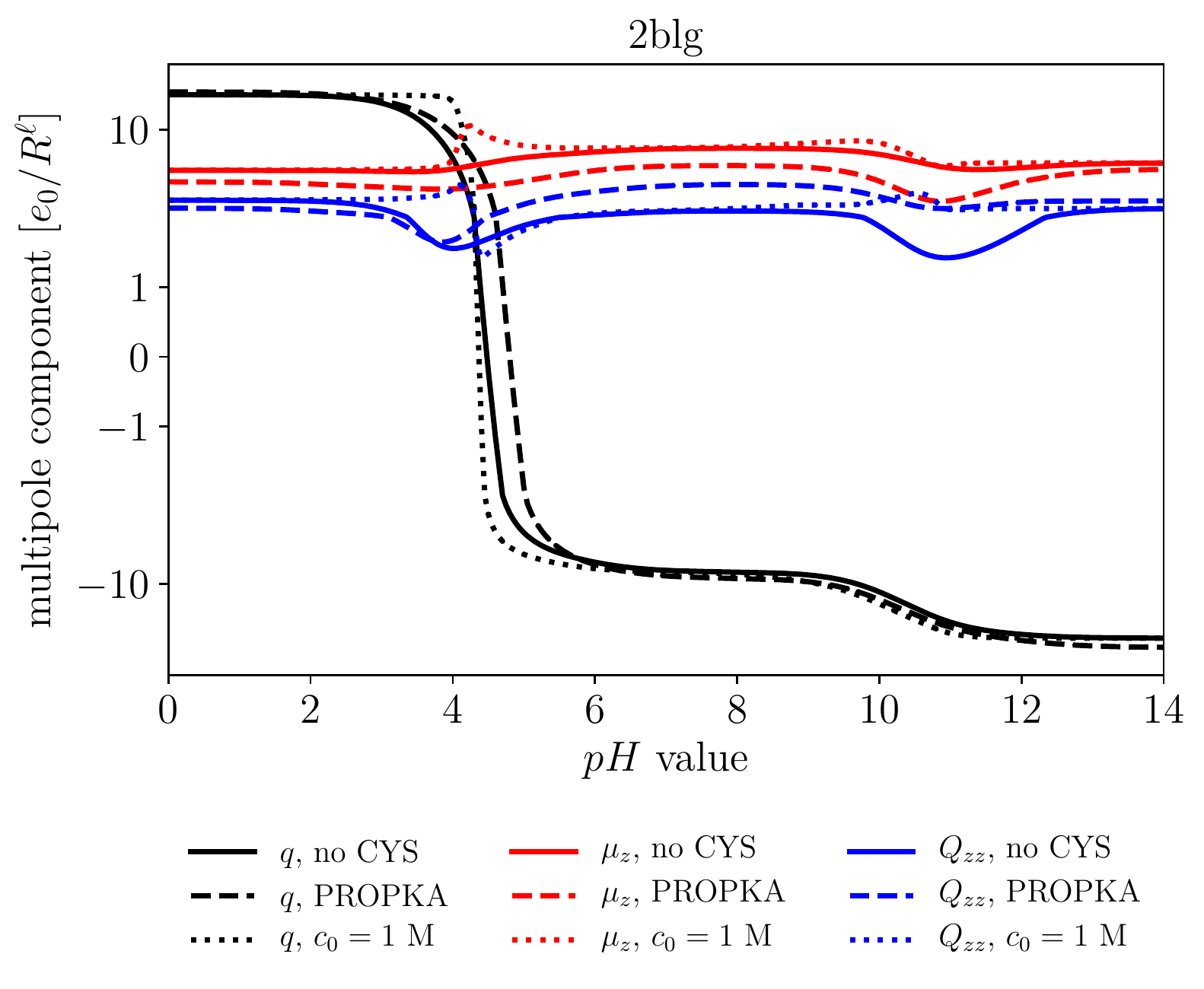}
\includegraphics[width=0.48\textwidth]{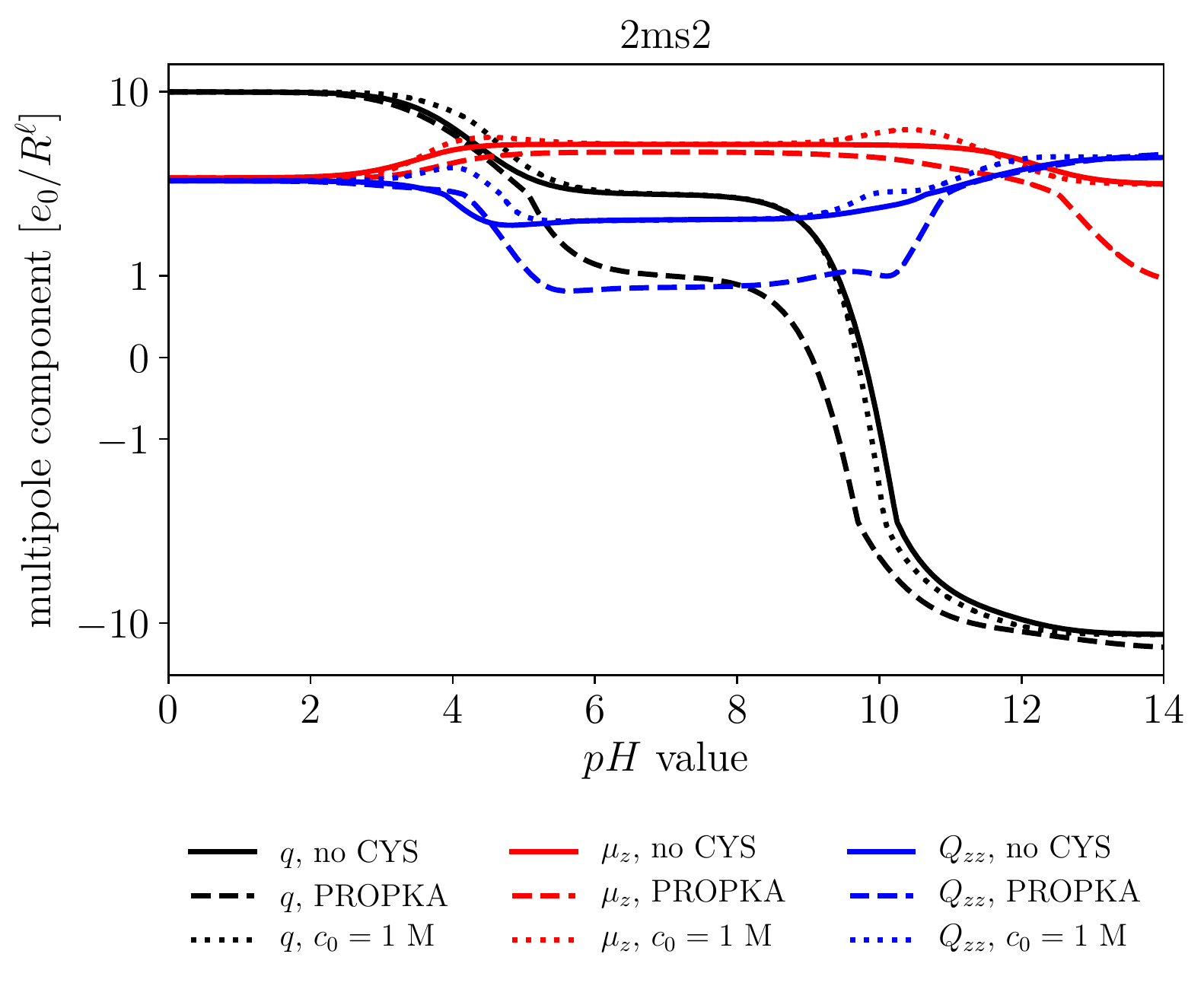}
\end{center}
\caption{Multipole components as a function of $pH$, comparing our simple model, where every amino acid group is assigned the same $pK_a$ value [cf.\ Table~\ref{tab:S2}], with two models introducing local effects to the $pK_a$ values of individual amino acid residues. From the latter two models, the first one uses empirical predictions of PROPKA software, which includes site-site interactions~\cite{Sondergaard2011}. The second model takes into account the local $pK_a$ shifts imparted by the electrostatic potential of the protein in the presence of 1:1 salt with bulk concentration $c_0=1$ M, derived using linearized DH equation with charge regulation boundary condition.  Shown for all four proteins: lysozyme (2lyz), human serum albumin (1e7h), $\beta$-lactoglobulin (2blg), and the MS2 capsid protein (2ms2). In our model, cysteine acidity is not considered.
\label{fig:S0}}
\end{figure*}

\clearpage
\newpage

\begin{figure*}[!htp]
\begin{center}
\includegraphics[width=0.325\textwidth]{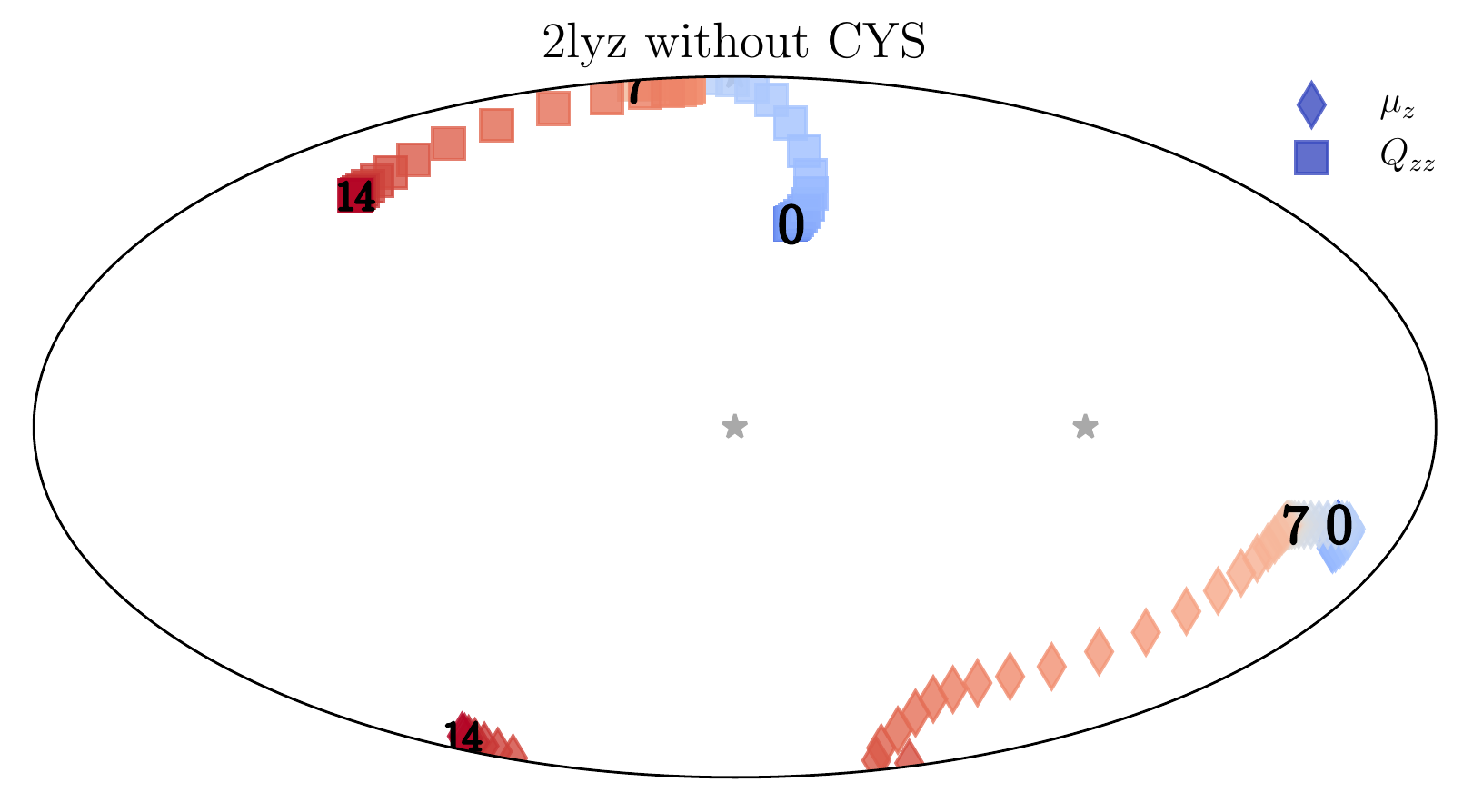}
\includegraphics[width=0.325\textwidth]{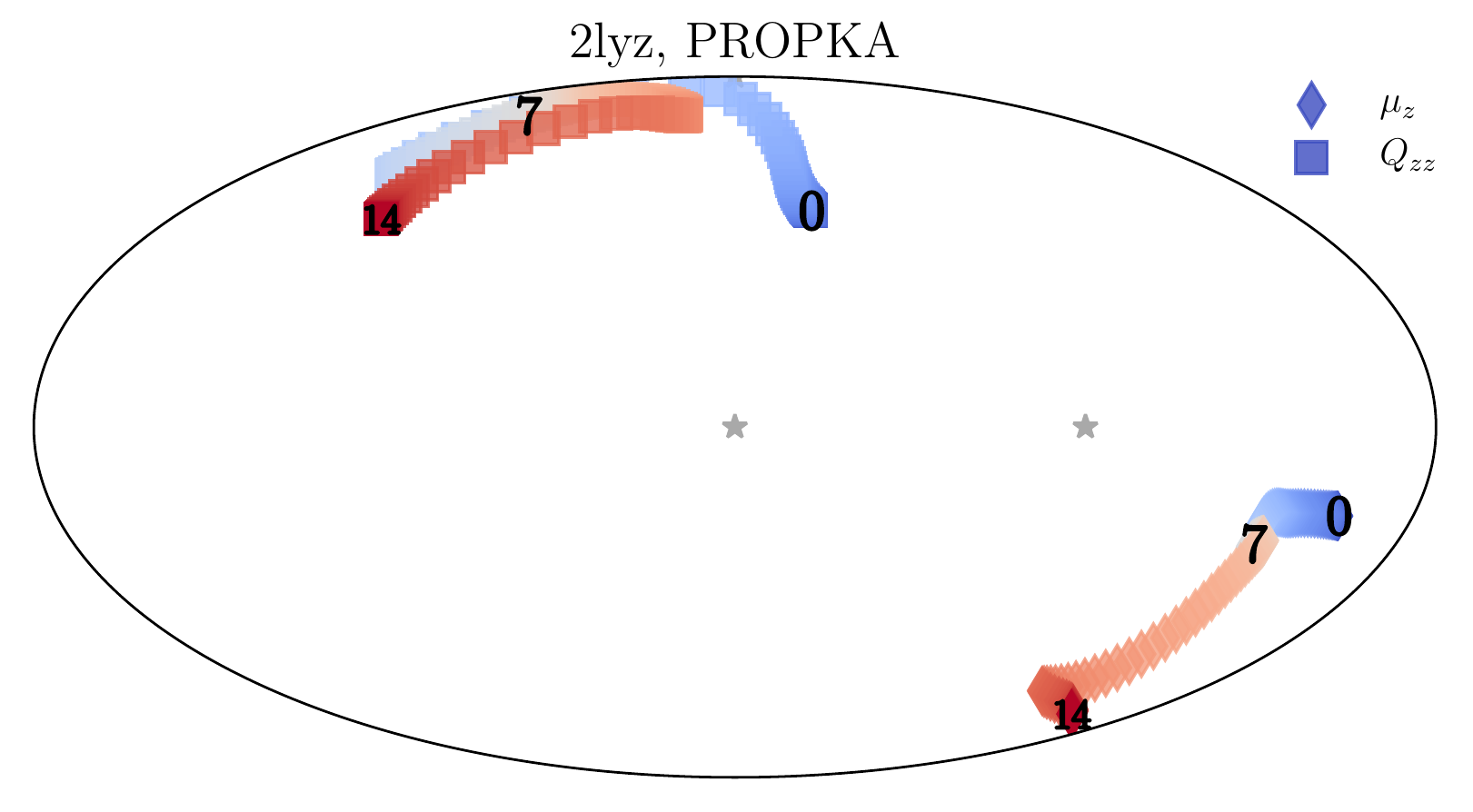}
\includegraphics[width=0.325\textwidth]{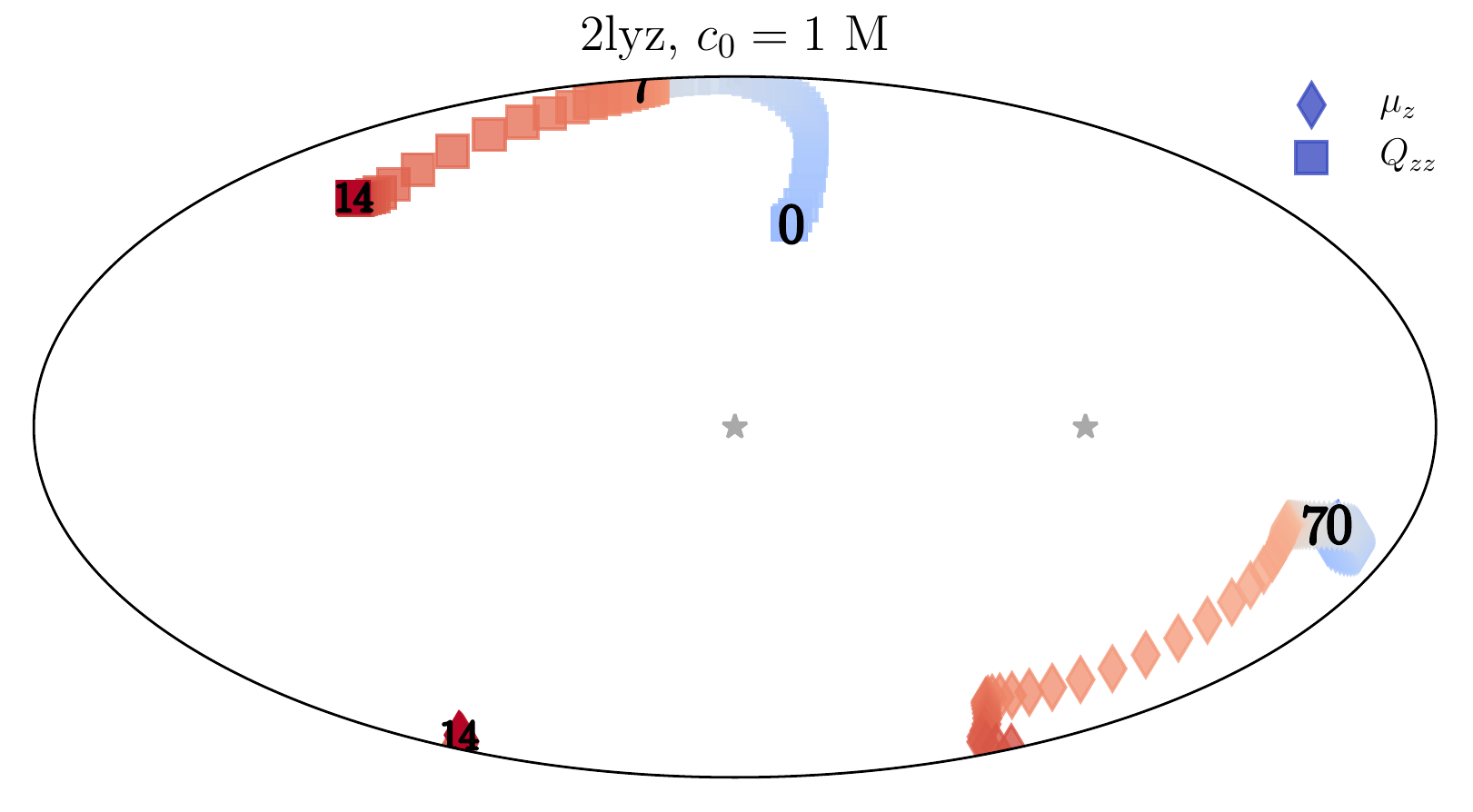}
\end{center}
\caption{Orientation of dipole and quadrupole principal $z$ axes of lysozyme (denoted by diamonds and squares, respectively) as a function of $pH$ (in steps of $0.2$ $pH$ unit). We compare our basic model, where the $pK_a$ values of ionizable amino acids are taken from Table~\ref{tab:S1}, with a model where the $pK_a$ values are empirically predicted by PROPKA based on site-site interactions, and with a model where the $pK_a$ values are shifted due to DH electrostatic potential in the presence of 1:1 salt with bulk concentration $c_0=1$ M. The $pH$ increase from 0 to 14 is shown with a color gradient, with blue hues denoting acidic $pH<7$ and red hues denoting basic $pH>7$. In addition, we explicitly indicate the positions at $pH$ values of 0, 7, and 14. The orientations of the axes are mapped from the circumscribed sphere of the protein onto a plane using the Mollweide projection; the gray stars show the coordinate axes of the original coordinate system of the protein, which is kept fixed.
\label{fig:S0a}}
\end{figure*}

\begin{figure*}[!htp]
\begin{center}
\includegraphics[width=0.325\textwidth]{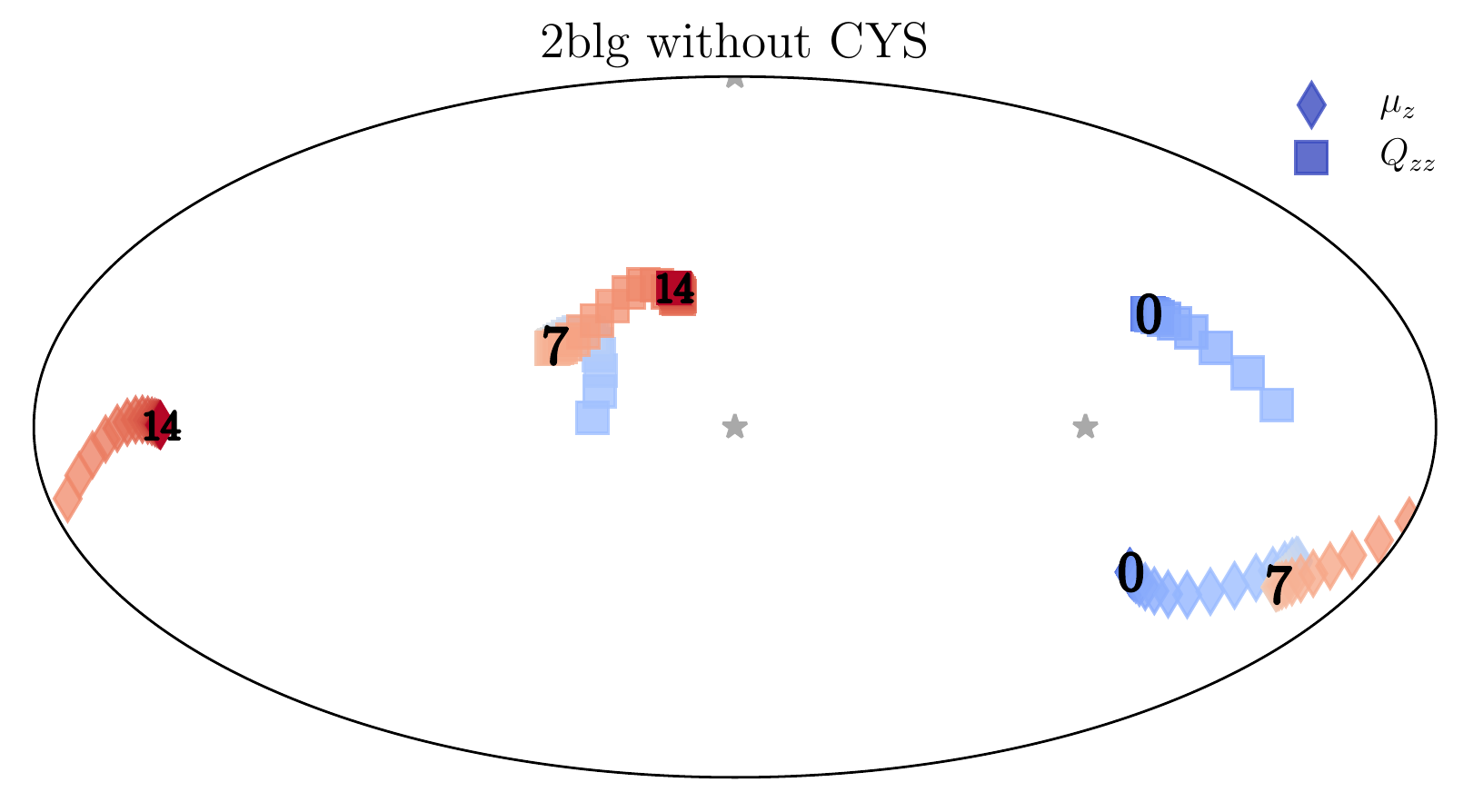}
\includegraphics[width=0.325\textwidth]{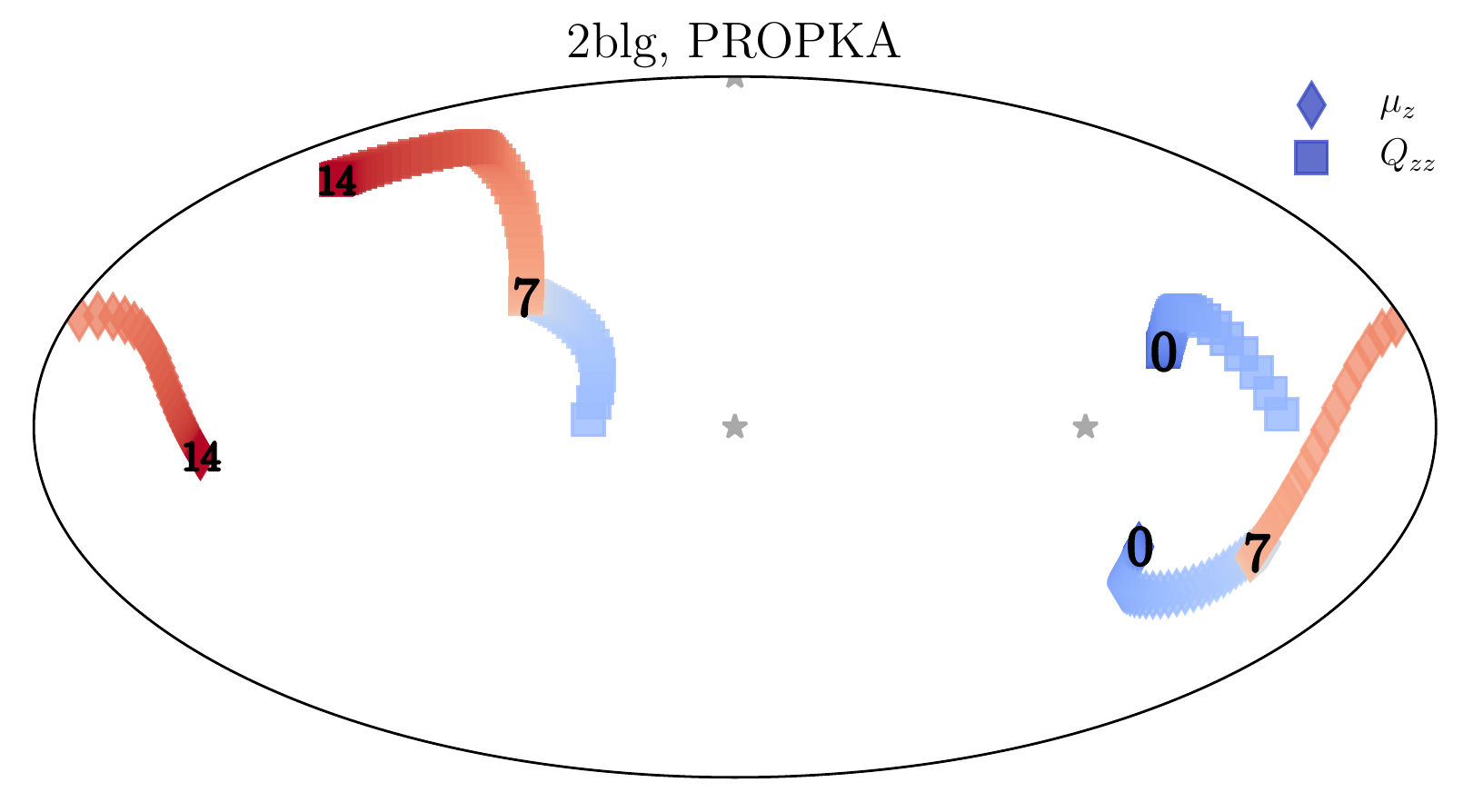}
\includegraphics[width=0.325\textwidth]{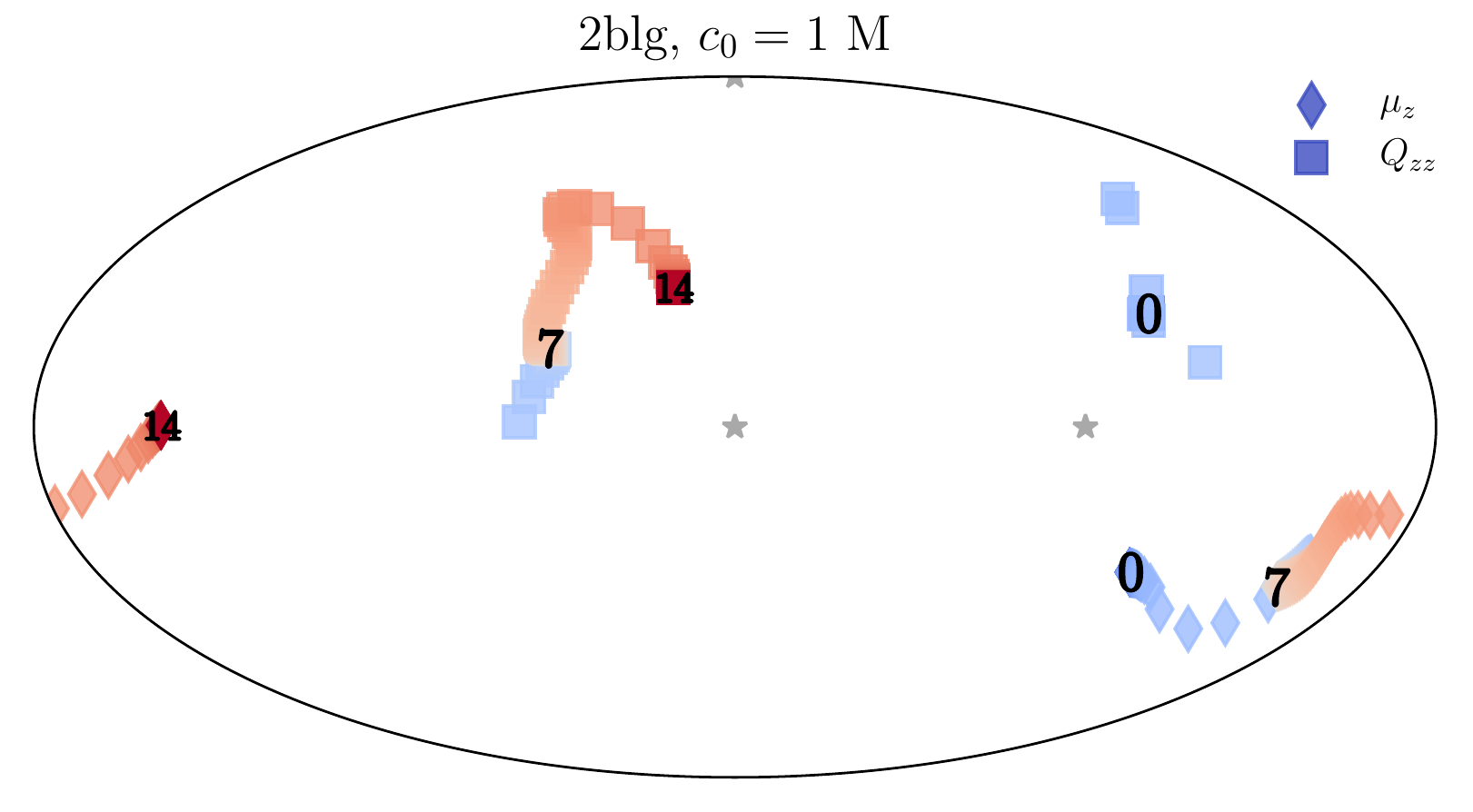}
\end{center}
\caption{Orientation of dipole and quadrupole principal $z$ axes of $\beta$-lactoglobulin (denoted by diamonds and squares, respectively) as a function of $pH$ (in steps of $0.2$ $pH$ unit). We compare our basic model, where the $pK_a$ values of ionizable amino acids are taken from Table~\ref{tab:S1}, with a model where the $pK_a$ values are empirically predicted by PROPKA based on site-site interactions, and with a model where the $pK_a$ values are shifted due to DH electrostatic potential in the presence of 1:1 salt with bulk concentration $c_0=1$ M. The $pH$ increase from 0 to 14 is shown with a color gradient, with blue hues denoting acidic $pH<7$ and red hues denoting basic $pH>7$. In addition, we explicitly indicate the positions at $pH$ values of 0, 7, and 14. The orientations of the axes are mapped from the circumscribed sphere of the protein onto a plane using the Mollweide projection; the gray stars show the coordinate axes of the original coordinate system of the protein, which is kept fixed.
\label{fig:S0b}}
\end{figure*}

\clearpage
\newpage

\section{EFFECT OF RSA CUT-OFF VARIATION}

\begin{figure*}[!htp]
\begin{center}
\includegraphics[width=0.48\textwidth]{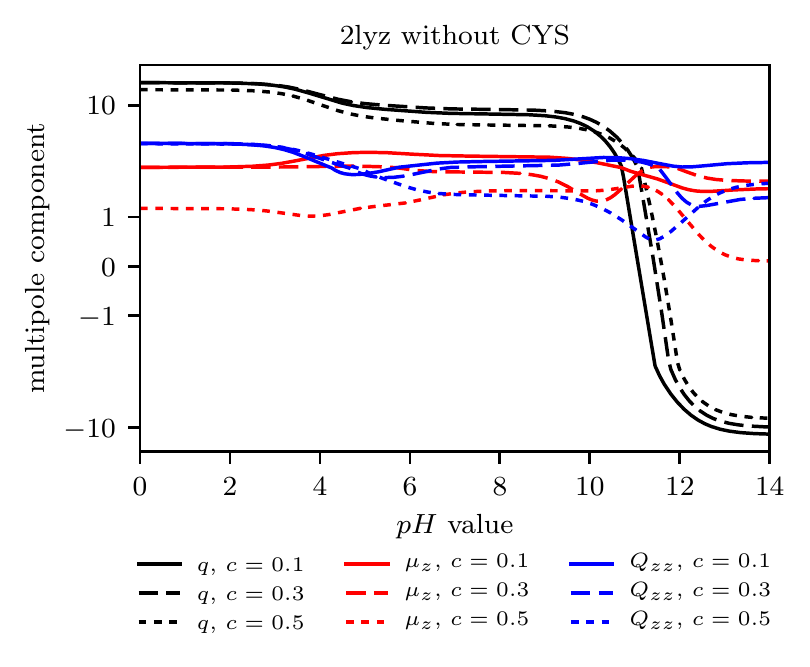}
\includegraphics[width=0.48\textwidth]{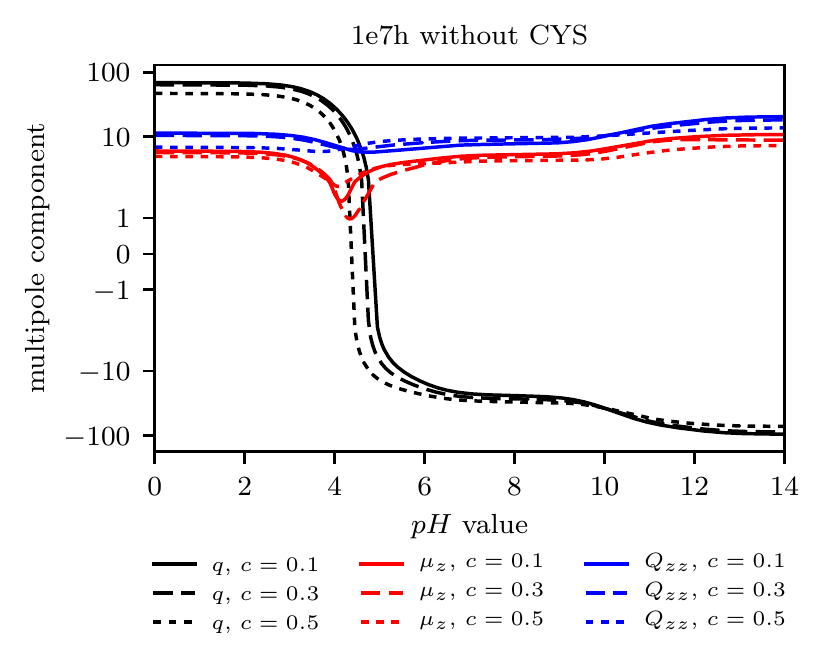}
\includegraphics[width=0.48\textwidth]{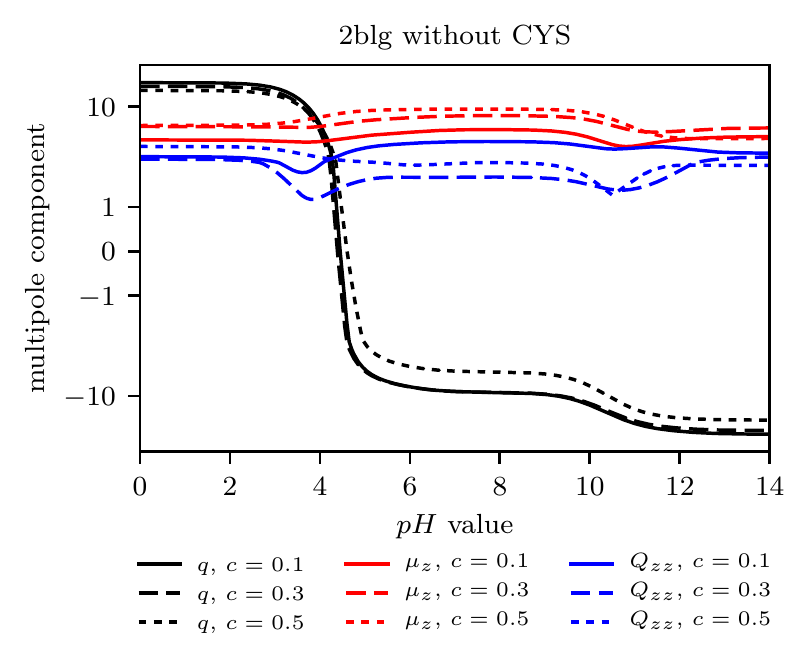}
\includegraphics[width=0.48\textwidth]{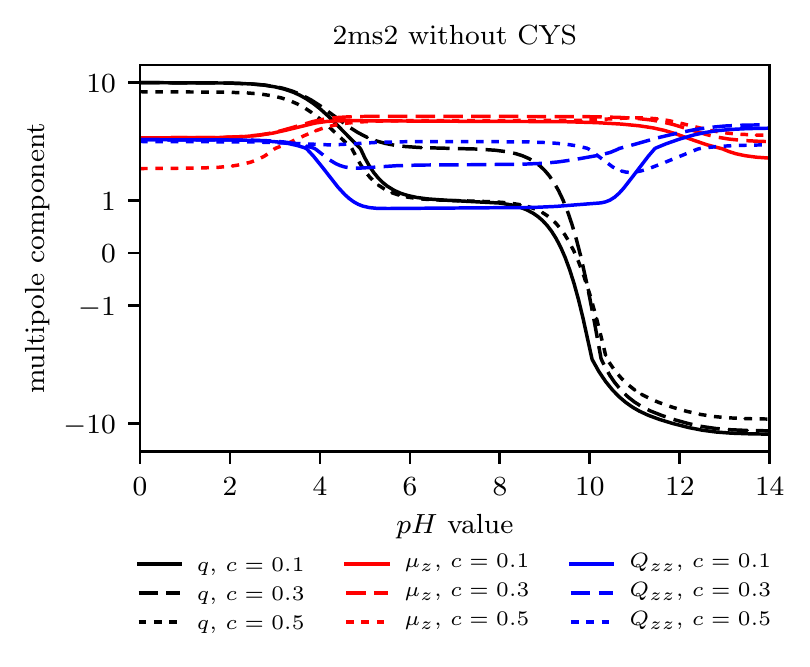}
\end{center}
\caption{Multipole components as a function of $pH$ for three different values of the RSA cut-off, $c=0.1$,$0.3$,$0.5$. Shown for all four proteins: lysozyme (2lyz), human serum albumin (1e7h), $\beta$-lactoglobulin (2blg), and the MS2 capsid protein (2ms2). Cysteine acidity is not considered.
\label{fig:S7}}
\end{figure*}

\clearpage
\newpage

\section{EFFECT OF CYSTEINE PROTONATION}

Cysteine has a thiol functional end group which is a very weak acid; in addition, it is reactive and can form disulfide bonds. For these reasons it is often not considered as an acid at all, and we thus did not take cysteine protonation into account in the main text. Here, we consider the effects in the case when the acidity of cysteine is considered. Cysteine residues comprise between $2$--$14$\% of our studied proteins' AA composition at the RSA cut-off of $c=0.25$ (cf.\ also Table~\ref{tab:S1}). Due to its $pK_a$ value (Table~\ref{tab:S2}), cysteine's region of influence is limited to $pH>7$ and its presence shifts charge distributions towards more negative values in that range~\citep{Nap2014}.

\begin{figure*}[!htp]
\begin{center}
\includegraphics[width=0.48\textwidth]{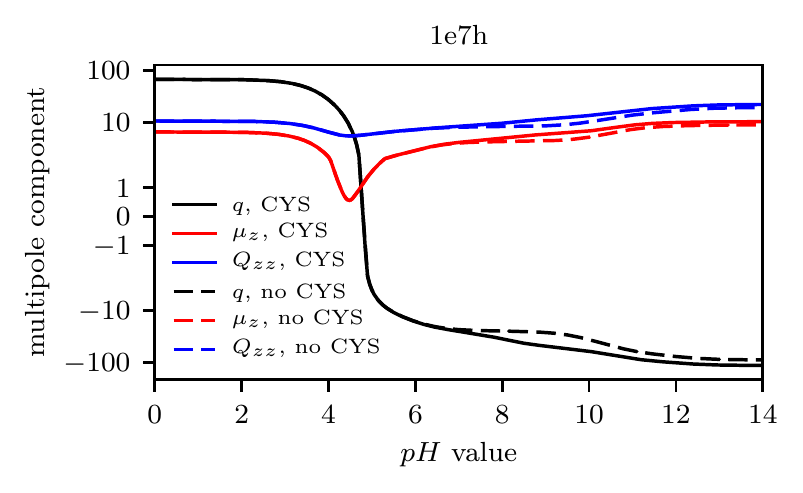}
\includegraphics[width=0.48\textwidth]{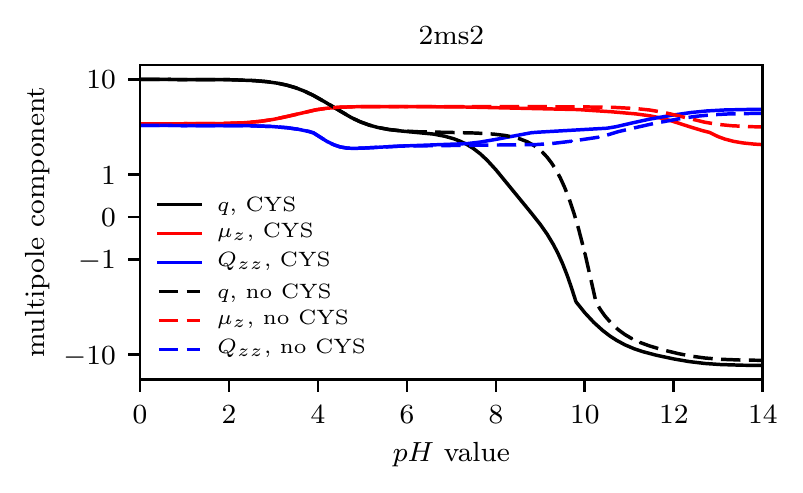}
\end{center}
\caption{Comparison of the magnitudes of the monopole ($q$), dipole ($\mu_z$), and quadrupole ($Q_{ii}$) components of the surface charge distribution as a function of $pH$ between the cases where cysteine acidity is considered (full lines) or is not considered (dashed lines). Shown for human serum albumin (1e7h) and phage MS2 capsid protein (2ms2).
\label{fig:S8}}
\end{figure*}

\begin{figure*}[!htp]
\begin{center}
\includegraphics[width=0.8\textwidth]{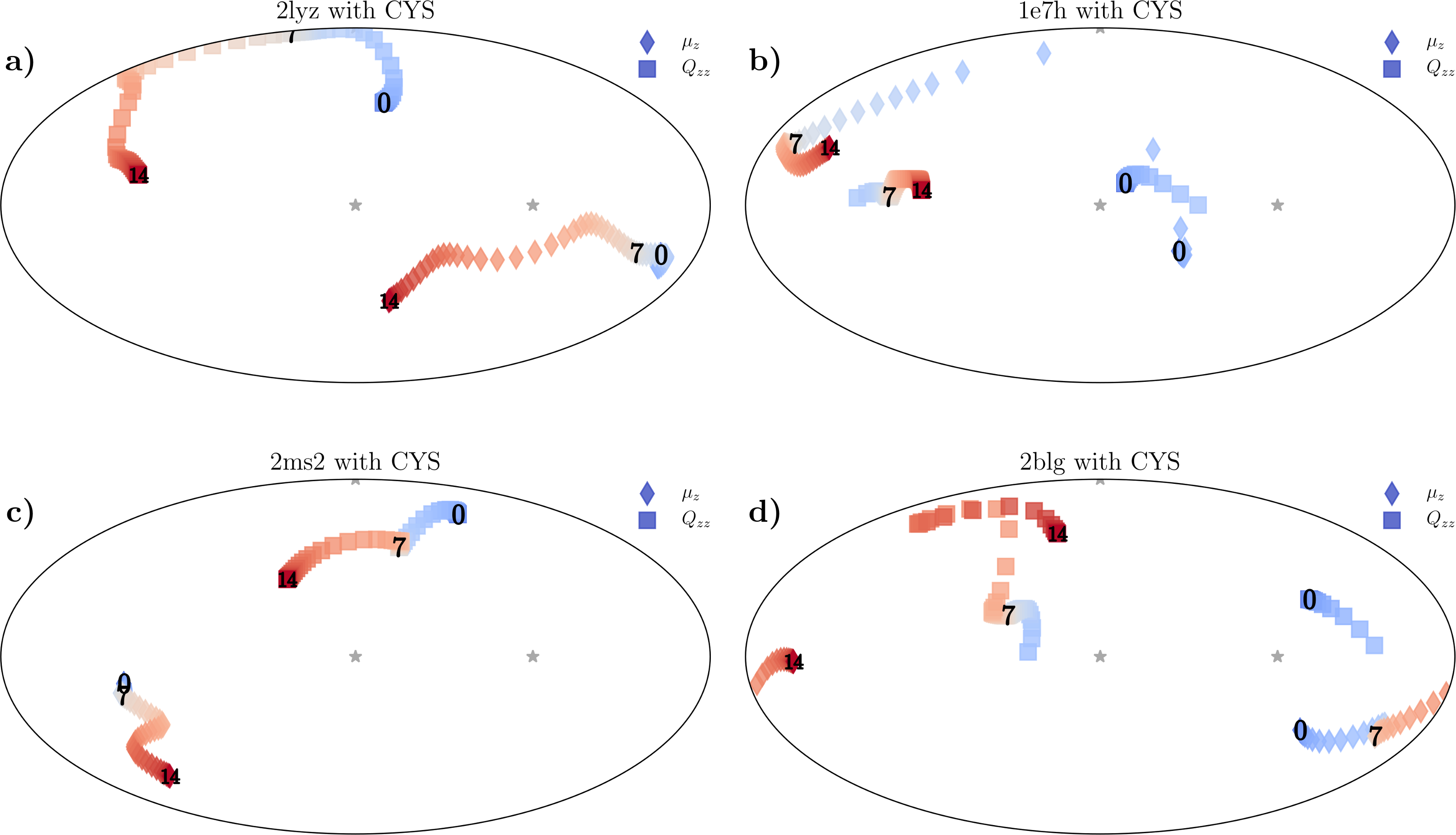}
\end{center}
\caption{{\bf (a)}-{\bf (d)} Orientation of dipole and quadrupole principal $z$ axes (denoted by diamonds and squares, respectively) as a function of $pH$ (in steps of $0.2$ $pH$ unit). The $pH$ increase from 0 to 14 is shown with a color gradient, with blue hues denoting acidic $pH<7$ and red hues denoting basic $pH>7$. In addition, we explicitly indicate the positions at $pH$ values of 0, 7, and 14. The orientations of the axes are mapped from the circumscribed sphere of the protein onto a plane using the Mollweide projection; the gray stars show the coordinate axes of the original coordinate system of the protein, which is kept fixed. The $pH$ dependence of the multipoles' orientation is shown side by side for all four proteins included in our study. Cysteine acidity is considered; compare with Fig.~6 in the main text.
\label{fig:S9}}
\end{figure*}

\end{document}